\documentclass[aps,reprint,superscriptaddress,nofootinbib]{revtex4-2}
\usepackage[utf8]{inputenc}
\usepackage{amsmath}
\usepackage{amssymb}
\usepackage{physics}
\usepackage{bbold}
\usepackage{graphicx}
\usepackage{multirow}
\usepackage{hyperref}
\hypersetup{colorlinks=true, allcolors=blue}
\usepackage{cleveref}
\usepackage{xcolor}
\usepackage{bbold} 
\usepackage[T1]{fontenc} 
\usepackage[caption = false]{subfig}

\usepackage{varwidth}
\usepackage{tcolorbox}

\usepackage{bm} 
\usepackage{MnSymbol} 

\definecolor{pink}{HTML}{F282B4}

\newcommand\D{\!\operatorname{d}\!}
\newcommand{\rh}[1]{\hat{\mathrm{#1}}}

\newcommand{\NC}{\mathcal{N}}
\newcommand{\EC}{\mathcal{E}}

\newcommand{\UC}{\mathcal{U}}
\newcommand{\Z}{\mathcal{Z}_2}
\newcommand{\cairo}{\textit{ibmq\_\hspace{-0.12cm}cairo}}
\newcommand{\cusco}{\textit{ibmq\_\hspace{-0.12cm}cusco}}

\usepackage{enumitem}

\newtheorem{definition}{Definition}
\newtheorem{lemma}{Lemma}

\begin{document}
\newcommand{\orcidicon}[1]{\href{https://orcid.org/#1}{\includegraphics[height=\fontcharht\font`\B]{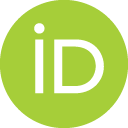}}}

\title{Symmetry breaking in geometric quantum machine learning in the presence of noise}


\author{Cenk T\"uys\"uz\orcidicon{0000-0003-0257-9784}}
\email{cenk.tueysuez@desy.de}
\affiliation{Deutsches Elektronen-Synchrotron DESY, 15738
Zeuthen, Germany}
\affiliation{Institut für Physik, Humboldt-Universit\"at zu Berlin, 12489 Berlin, Germany}

\author{Su Yeon Chang\orcidicon{0000-0001-5768-2434}}
\affiliation{European Organization for Nuclear Research (CERN), 1211 Geneva, Switzerland}
\affiliation{Institute of Physics, \'Ecole Polytechnique F\'ed\'erale de Lausanne (EPFL), 1015 Lausanne, Switzerland }

\author{Maria Demidik}
\affiliation{Deutsches Elektronen-Synchrotron DESY, 15738
Zeuthen, Germany}
\affiliation{Computation-Based Science and Technology Research Center, The Cyprus Institute, 2121 Nicosia, Cyprus}

\author{Karl Jansen}
\affiliation{Deutsches Elektronen-Synchrotron DESY, 15738
Zeuthen, Germany}
\affiliation{Computation-Based Science and Technology Research Center, The Cyprus Institute, 2121 Nicosia, Cyprus}

\author{Sofia Vallecorsa}
\affiliation{European Organization for Nuclear Research (CERN), 1211 Geneva, Switzerland}

\author{Michele Grossi\orcidicon{0000-0003-1718-1314}}
\email{michele.grossi@cern.ch}
\affiliation{European Organization for Nuclear Research (CERN), 1211 Geneva, Switzerland}

\begin{abstract}
    Geometric quantum machine learning based on equivariant quantum neural networks (EQNN) recently appeared as a promising direction in quantum machine learning. Despite the encouraging progress, the studies are still limited to theory, and the role of hardware noise in EQNN training has never been explored. This work studies the behavior of EQNN models in the presence of noise. We show that certain EQNN models can preserve equivariance under Pauli channels, while this is not possible under the amplitude damping channel. We claim that the symmetry breaking grows linearly in the number of layers and noise strength. We support our claims with numerical data from simulations as well as hardware up to 64 qubits. Furthermore, we provide strategies to enhance the symmetry protection of EQNN models in the presence of noise.
\end{abstract}

\maketitle

\section{Introduction}

Variational quantum algorithms (VQAs) appear to be one of the promising algorithms of the noisy intermediate scale quantum (NISQ) era~\cite{Preskill2018} in the literature~\cite{cerezo_variational_2021}. Furthermore, recent results showed noise resilience of VQAs, which further increased hope~\cite{fontana_evaluating_2021}. However, there exist many roadblocks to making this promise a reality. Some problems that are common to most VQAs are barren plateaus (BPs) \textit{i.e.} number of shots needed to estimate the sufficiently precise values of the cost function grows exponentially~\cite{McClean2018BP, ragone2023unified}, many local minima~\cite{anschuetz_quantum_2022, You2023minima, Rivera2021minima} and lack of efficient gradient computation (\textit{e.g.} parameter shift rules require circuit executions that scale linearly in number of parameters)~\cite{wierichs_general_2022}. While certain issues can be partially alleviated through a range of methods~\cite{Grant2019, volkoff_large_2021, sack_avoiding_2022, patti_entanglement_2021, tuysuz_classical_2023}, faithfully running these algorithms on NISQ hardware, beyond what is classically simulable (\textit{e.g.} $n > 40$ qubits and at least log$(n)$ depth), is still a practical challenge.

Proposals of geometric quantum machine learning (GQML) opened new avenues, which in theory bring VQAs closer to practicality~\cite{Larocca2022GQML}. The GQML framework leverages inductive biases on problems and uses this to construct algorithms with improved trainability and generalization~\cite{Meyer2023Symmetry}. This requires the circuit to have a certain structure from the initial state until the final measurements. On the other hand, this is where the NISQ hardware fails to provide due to coherent and incoherent errors present~\cite{Preskill2018, bharti2022noisy}. In the literature, this topic has been explored in the context of state preparation and time evolution of quantum systems, in which many physical symmetries arise~\cite{tran_faster_2021, nguyen_digital_2022}. However, these results don't directly translate to the setting of GQML. For this reason, we study the behavior of these algorithms, specifically equivariant quantum neural networks (EQNNs), under hardware noise in this work.

In this paper, we study the behavior of EQNN models in the presence of noise. Our theoretical and numerical results indicate that, for the models considered, equivariance can be protected under realistic Pauli channels. We further show that the symmetry is broken under the non-unital amplitude damping channel. We characterize this with metrics that we introduce and show that symmetry breaking grows approximately linearly in the number of layers and the noise strength. Moreover, we provide strategies such as choice of representation and adaptive thresholding to improve performance.


We structure the paper as follows. In Section~\ref{sec:framework}, we introduce the necessary preliminary definitions from how to construct EQNNs, to how the hardware noise is modeled. Then, in Section~\ref{sec:theory}, we start by constructing a toy model and use it to show how hardware noise can break the equivariance. After establishing the theoretical intuition, we define data-driven metrics to quantify the symmetry breaking. Section~\ref{sec:experiments} consists of numerical experiments performed with classical simulators as well as NISQ hardware. In Section~\ref{sec:conclusion}, we share our point of view on what error mitigation means for the results that we establish, and we conclude by giving suggestions on deploying EQNN models on hardware and talk about future directions and some open questions.

\section{Framework}
\label{sec:framework}

\subsection{Equivariant Quantum Neural Networks}

This paper focuses on the supervised learning task over a classical data space $\mathcal{R}$, where the data point $\mathbf{x}_i \in \mathcal{R}$ is associated with a label $y_i\in\mathcal{Y}$ following the hidden distribution $f: \mathcal{R}\to \mathcal{Y}$.
In the most general framework of quantum machine learning (QML) manipulating the classical data, we embed each $\mathbf{x}_i$ into a quantum state $\rho_{\mathbf{x}_i}\in \mathcal{M}$ with a certain quantum feature map $\Psi:\mathcal{R} \to \mathcal{M}$ where $\mathcal{M}$ is the space of semidefinite positive density matrices~\cite{havlivcek2019supervised}. The input state is transformed via a quantum map $\mathcal{U}_{\bm{\theta}}(\rho)$, which is the adjoint action of $U_{\bm{\theta}}$ on state $\rho$, 
\begin{equation}
    \mathcal{U}_{\bm{\theta}}(\rho_{\mathbf{x}_i}) = U_{\bm{\theta}}   \rho_{\mathbf{x}_i} U_{\bm{\theta}}^\dagger 
\end{equation}  with $U_{\bm{\theta}}$ the quantum neural network (QNN) is parameterized by a set of trainable parameters $\bm{\theta}$. 
Without losing generality, we consider the most general setup where the final prediction of the QNN is the expectation value of an observable $O$ : 
\begin{equation}
    \hat{y}(\mathbf{x}) = \hat{f}_{\bm{\theta}}(\rho_{\mathbf{x}}) = {\rm Tr}[\mathcal{U}_\theta(\rho_{\mathbf{x}}) O].
\end{equation}
During the training, the model learns the hidden data distribution from the training set in such way that $\hat{f}_{\bm{\theta}}$ approaches as close as possible to the target function $f$.  At the end of the training, we expect that $\mathcal{U}_{\bm{\theta}}$ can also predict the label of the unseen test set. 

The key idea behind geometric quantum machine learning (GQML) is to design models that capture the meaningful relations in the dataset by incorporating the architecture with the geometric priors. In the case of geometric supervised learning, we consider the \textit{label symmetry} of the dataset given as the following definition.

\begin{definition}[Invariance]
    Let us consider a symmetry group $\mathcal{S}$ with a representation $R :\mathcal{S} \to {\rm Aut}(\mathcal{R})$ acting on the classical data space $\mathcal{R}$. We call that a function $h$ has a label symmetry if and only if $h$ is \textit{invariant} under $\mathcal{S}$, i.e.,
\begin{equation}
    h(\rho_{R(g)\cdot \mathbf{x}}) = h(\rho_{\mathbf{x}}),~\forall g \in \mathcal{S}.
\end{equation}
\end{definition}


GQML aims to construct a QNN ansatz that guarantees this label symmetry so that the final prediction $\hat{y}(\mathbf{x})$ is invariant under the action of any symmetry group element $g\in \mathcal{S}$. Recent papers suggest approaching the GQML with $\mathcal{S}$-\textit{equivariant quantum model}~\cite{Meyer2023Symmetry, Larocca2022GQML}.

\begin{definition}[Equivariant Embedding]
    We call an embedding $\Psi: \mathcal{R}\to \mathcal{M}$ with $\Psi(\mathbf{x}) = \rho_{\mathbf{x}}$ to be \textbf{equivariant} with respect to a symmetry element $g$, if and only if there exist a unitary representation $R_q(g)$ such that  
\begin{equation}
    \rho_{R(g) \cdot \mathbf{x}} =  R_q(g) \rho_{\mathbf{x}} R_q^\dagger(g).
\end{equation}
\end{definition}
We call $R_q(g)$ the unitary representation of $g$ \textit{induced} by the embedding $\Psi$~\cite{Meyer2023Symmetry}. 
The group symmetry emerges naturally in the QNN architecture via the equivariant embedding and can be captured by the equivariant quantum gates. For simplicity, let us focus on a set of quantum gates of the form : 
\begin{equation}
    U_G(\theta) = \exp(-i\theta G),~~G \in \mathcal{G}
\end{equation}
where $G$ is a Hermitian generator and $\mathcal{G}$ the generator set of $U$. 

\begin{definition}[Equivariant Gate]
    A quantum gate $U_G(\theta) = \exp(-i\theta G)$ with $\theta \in \mathbb{R}$ is called to be \textbf{equivariant} with respect to $\mathcal{S}$ if and only if it commutes with $R_q(g)$ for all $g\in\mathcal{S}$, i.e.,  

    \begin{equation}
      [U_G(\theta), R_q(g)] = 0,~~\forall \theta \in \mathbb{R}, \forall g \in  \mathcal{S} 
    \end{equation}
or equivalently,  
\begin{equation}
    [G, R_q(g)] = 0,~~\forall g \in  \mathcal{S}. 
\end{equation}
\end{definition}
There exist different methods proposed to construct the equivariant gateset~\cite{nguyen2022theory}, such as \textit{twirling} method, which is the most common and practical method for a finite symmetry group. 

Similarly, a QNN ansatz is said to be equivariant if and only if it consists of equivariant quantum gates. By combining the equivariant embedding and the equivariant QNN ansatz with an equivariant observable $O$ : 
\begin{equation}
    R_q(g) O R_q^\dagger(g) = O,~~\forall g \in \mathcal{S}, 
\end{equation}
we construct an \textit{\bf invariant quantum classifier model} which guarantees this label symmetry. 

\begin{lemma}[Invariance from equivariance] A quantum learning model which consists of equivariant embedding, equivariant quantum circuit ansatz and invariant observable with respect to a symmetry group $\mathcal{S}$ is invariant with respect to $\mathcal{S}$  : 
\begin{align}
    \hat{y}(R(g) \cdot \mathbf{x}) & = {\rm Tr}[U_{\bm{\theta}} \rho_{R(g) \cdot \mathbf{x}}   U^\dagger_{\bm{\theta}}  O]  \nonumber  \\ & = {\rm Tr}[U_{\bm{\theta}} R_q(g)\rho_{\mathbf{x}}  R_q^\dagger(g) U^\dagger_{\bm{\theta}}  O]  \nonumber  \\  & =  {\rm Tr}[R_q(g) U_{\bm{\theta}}  \rho_{\mathbf{x}} U^\dagger_{\bm{\theta}} O  R_q^\dagger(g)] \nonumber  \\  & = {\rm Tr}[R_q^\dagger(g)  R_q(g) \mathcal{U}_\theta(\rho_{\mathbf{x}}) O ] \nonumber  \\  & = {\rm Tr}[\mathcal{U}_\theta(\rho_{\mathbf{x}})  O ] = \hat{y}(\mathbf{x}),~~\forall g\in \mathcal{S}. 
\end{align}
\end{lemma}

The equivariant QNN leads to the trade-off between the gain of expressibility and the loss of expressibility by constraining the search space that the model can explore. 
In the previous studies, GQML has shown promising results in various problem setups leveraging the advantage in terms of complexity, trainability and 
generalization~\cite{kazi2023GQML,Larocca2022GQML,nguyen2022theory,Ragone2023GQML,schatzki2022theoretical,Zheng2023Equiv, chang2023approximately}. However, all the tests have been undertaken in the absence of hardware noise and the impact of noise on EQNN has never been studied before. 

\subsection{Noise models}
\label{sec:noise}
The description of noise effects during quantum gates operation is based on the open quantum system theory~\cite{breuer2002theory, NielsenChuang2010}. 
The Markovian evolution of the density matrix $\rh{\rho}_t$ of the qubits system in a given environment is described by the following Lindblad equation
\begin{equation}
\label{mastereq}
\frac{\D}{\D t} \rh{\rho}_{t} = -\frac{i}{\hbar}[\rh{H}_{t},\hat{\rho}_{t}] + \mathcal{L}\rh{\rho}_{t},
\end{equation}
where
\begin{equation}
\mathcal{L}\rh{\rho}_{t}=\epsilon^{2} \sum_{k}\biggl[\rh{L}_{k}\rh{\rho}_{t}\rh{L^{\dagger}}_{k} - \frac{1}{2}\{\rh{L^{\dagger}}_{k}\rh{L}_{k},\rh{\rho}_{t}\}\biggr],
\end{equation}
$\rh{H}_{t}$ is the time-dependent Hamiltonian realizing a given gate, and $\rh{L}_{k}$ are the Lindblad operators capturing the action of the environment.

In this work, we only consider quantum channels acting locally on qubits. Some examples of these channels are \textit{bit flip} (BF) channel, \textit{depolarizing} (DP) channel, and \textit{amplitude damping} (AD) channel. One way to define the action of a noise channel $\NC$ on the quantum state $\rho$ is through the Kraus operators $K$~\cite{NielsenChuang2010}. Then this can be written as,

\begin{equation}
    \NC(\rho) = \sum_i K_i \rho K_i^\dagger.
\end{equation}

\begin{description}[leftmargin = 0pt]

\item[Bit Flip Channel] BF channel with probability $p$ can be described using two Kraus operators $K_0 = \sqrt{1-p}~I$ and $K_1 = \sqrt{p}~X$. The action of the BF channel on the single qubit state simply becomes,

\begin{equation}
    \NC(\rho) = (1-p)\rho + pX \rho X. 
\end{equation}

This can be extended to multi-qubit systems. In the two-qubit case, the action of the noise channel can be written as, 

\begin{align}
    \NC(\rho) =&~(1-p_0)(1-p_1)\rho \nonumber \\
    &+ p_0(1-p_1)(X \otimes I) \rho (X \otimes I) \nonumber \\
    &+ (1-p_0)p_1 (I \otimes X) \rho (I \otimes X) \nonumber \\
    &+ p_0p_1(X \otimes X) \rho (X \otimes X), 
    \label{eq:2q-bitflip}
\end{align}

where $p_0$ and $p_1$ are the probability of acting on qubit-0 and qubit-1, respectively. Following this logic, all local noise channels can be generalized to multi-qubit systems. 

\item[Depolarizing Channel] Kraus operators of the DP channel are $K_0 = \sqrt{1-p}~I$, $K_1 = \sqrt{p/3}~X$, $K_2 = \sqrt{p/3}~Y$, $K_3 = \sqrt{p/3}~Z$. Single qubit DP channel shrinks the Bloch sphere from all directions symmetrically. Hence, any quantum state moves towards the maximally mixed state under the action DP channel. 

\item[Pauli Channel] Both BF and DP channel are special cases of Pauli channels. Kraus operators of the Pauli channel are $K_0 = \sqrt{1-p_x - p_y - p_z}~I$, $K_1 = \sqrt{p_x/3}~X$, $K_2 = \sqrt{p_y/3}~Y$, $K_3 = \sqrt{p_z/3}~Z$. One can recover the BF channel by setting $p_y=p_z=0$ and the DP channel by setting $p_x=p_y=p_z=p$. 

\item[Amplitude Damping Channel] The picture changes significantly under AD channel. Kraus operators of the AD channel can be written as,

\begin{equation}
K_0 = \begin{bmatrix}
    1& 0\\
    0& \sqrt{1-\gamma}
\end{bmatrix}, \,
K_1 = \begin{bmatrix}
    0& \sqrt{\gamma}\\
    0& 0
\end{bmatrix}, 
\end{equation}
with $\gamma$ the amplitude damping probability.

The action of single qubit AD channel shrinks the Bloch sphere towards the ground state ($\ket{0}$), creating an asymmetry on the Hilbert space along the $z$-direction. Another common way of describing the noise channels is through the \textit{Pauli transfer matrix} (PTM) formalism~\cite{ptm}. This simplifies some computations and is used in this work. Please refer to Appendix~\ref{app:ptm} for more details on the PTM formalism. 
\end{description}
Having these definitions, we can now describe the action of noise on the quantum circuit. Let us consider a quantum system with initial state $\rho_0$ and at every layer the circuit acts with the unitary $U_i$, such that $\rho_i = \UC_i\left( \rho_{i-1} \right) = U_i\rho_{i-1}U_i^\dagger$. Then, the quantum state, after layer $d$ becomes,

\begin{equation}
    \rho_d = \NC \circ \UC_d \circ \cdots \circ \NC \circ \UC_2 \circ \NC \circ \UC_1(\rho_0).
\end{equation}

\begin{figure}[!h]
    \centering
    \includegraphics[width=.8\linewidth]{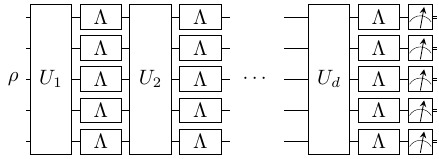}
    \caption{Drawing of the local noise model. A circuit with input $\rho$ and layers $U_i$, where the local $\Lambda$ representing the action of noise are applied after each layer.}
    \label{fig:noise-model}
\end{figure}

This can be visualized in the circuit picture as in Fig.~\ref{fig:noise-model}, where $\Lambda$ is the local action of noise channel $\NC$. On the real hardware, the action of $\Lambda$ is different for all qubits, but for simplicity, we assume that they are the same for simulations.

An extended description of the noise channels can be found in Appendix~\ref{app:noise}.

\subsection{Concentration of measure}

Variational algorithms may experience an exponential concentration of measure mainly due to the fact that the quantum state living in the $2^n$-dimensional Hilbert space. The term \textit{concentration of measure} refers to the observation that in many high-dimensional spaces, continuous functions are almost everywhere close to their mean~\cite{muller2011concentration}. As defined in Definition~\ref{def:exp-conc}, the exponential concentration is commonly referred to as barren plateaus (BPs) in the QML literature. It was shown on different occasions that BPs may exist due to excessive expressivity of the circuit~\cite{holmes_connecting_2022}, highly entangled input state (\textit{e.g.} volume law entanglement)~\cite{ortiz_marrero_entanglement-induced_2021}, global observables~\cite{Cerezo2021CostBP} and hardware noise~\cite{wang_noise-induced_2021}. We refer the readers to recent work by Ragone et al.~\cite{ragone2023unified}, which offers a unified picture of these causes.

\begin{definition}[Exponential concentration] Consider the random variable $X$. X is said to be deterministically exponentially concentrated in the number of qubits $n$ around a certain fixed value $\alpha$ for some $b > 1$ if
\begin{equation}
    \abs{X-\alpha} \leq \beta \in \mathcal{O}(1/b^n).
\end{equation}
\label{def:exp-conc}
\end{definition}

EQNNs can avoid BPs\footnote{Here, the term avoiding barren plateaus, is used in the context of barren plateaus, where the concentration of measure is caused by the \textit{expressivity} of the ansatz. This doesn't hold for all cases, \textit{e.g.} global observables, noise-induced BPs etc. } by incorporating inductive biases into the ansatz design. This follows the fact that the existence of BPs in the case where the ansatz admits a Lie algebra $\mathfrak{g}$, such that dim$(\mathfrak{g}) \in \mathcal{O}(\mbox{exp}(n))$. Consequently, certain EQNNs can be constructed in a polynomial subspace of the Hilbert space such that they admit dim$(\mathfrak{g}) \in \mathcal{O}(\mbox{poly}(n))$ and allow BP-free parametrized circuit designs~\cite{ragone2023unified}. This framework can ensure BP-free models unless there is no hardware noise present. Inevitably, EQNNs will experience noise-induced BPs~\cite{wang_noise-induced_2021}. This will be an important point when discussing the performance of EQNNs in the presence of noise.

\section{Equivariance under noise}
\label{sec:theory}

Writing down analytical expressions for noisy quantum circuits is a difficult task in general. The expressions are unique to each circuit and noise model, resulting in complicated equations with just a few layers of gates. Nonetheless, this offers a good understanding of the behavior of the model in simple settings. To be able to do this, we construct a toy model. This allows us to build a theoretical understanding and give us an intuition of what to expect from numerical results. 

\subsection{Toy model}
\label{sec:toy}

\begin{figure}[!h]
    \centering
    \includegraphics{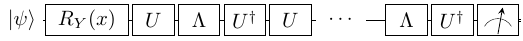}
    \caption{One qubit toy model under noise with identity gates decomposed into unitaries $U$ and $U^\dagger$, $d$ times, i.e., $I = (UU^\dagger)^d$.}
    \label{fig:1q-toy}
\end{figure}

Let us consider the following circuit, where the one-dimensional input data $x \in \mathbb{R}$ is encoded using the $R_Y$ rotation gate. Then, we will apply an identity gate that we decompose into the form $U U^\dagger$, $d$ times. This formulation will allow us to incorporate the effects of gate decompositions on the behavior of the circuit. When designing algorithms in the NISQ era, we should keep in mind that we only have access to a limited set of native gates on hardware.  The noise channel $\Lambda$ will be applied between each $U$ and $U^\dagger$ gates as described in Fig.~\ref{fig:1q-toy}.

We assume a dataset with the $\Z=\{e,\sigma\}$ symmetry, such that $R(e) \cdot x=x$ and $R(\sigma) \cdot x =-x$. Then, one can use any rotation gate $R_G$, such that the twirl with the representation $R_q(\sigma)$ is $R_q(\sigma) G R_q^\dagger(\sigma) = -G$. Then, one can use the $R_Y$ rotation gate to encode this symmetry simply due to the fact that $XYX=-Y$ and similarly $ZYZ=-Y$. This means we have the freedom of choosing either $X$ or $Z$ as the representation $R_q(\sigma)$. The choice of representation is going to put a constraint also on the input state. For this walk-through, let's choose the input state $\ket{\psi} = \ket{+}$, $R_q(\sigma)=X$, $U=R_Y(\theta)$ and the choice of representation requires us to have the observable $O=X$. Here, we refer the reader to Refs.~\cite{Larocca2022GQML, Meyer2023Symmetry, nguyen2022theory} for more details on constructing EQNNs.

Having defined our complete model, we can now choose a noise model and express the model outputs analytically. We refer the reader to Appendix~\ref{app:toy} for step-by-step calculations in this section. First, we consider the Pauli channel. Then, the output of the model can be written as,

\begin{equation}
\label{eq:toy-pauli}
\hat{y}(x) = \frac{1}{4}\big((f_x^d+f_z^d)\cos(x)+(f_x^d-f_y^d)\cos(x+2\theta)\big),
\end{equation}

where $f_i$ is the Pauli fidelity of the Pauli $\sigma_i$ (\textit{e.g.} $f_x = 1-2(p_y+p_z)$ according to the definition in Section~\ref{sec:noise}, refer to Appendix~\ref{app:noise} for more details.). 
The first term of the equation gives us the noiseless outcome that is suppressed exponentially in the number of layers around zero (\textit{i.e.} $ \{|f_x|, |f_y|, |f_z| \} \leq 1$). This result is also known as noise-induced barren plateaus~\cite{wang_noise-induced_2021}. The second term of the equation constitutes the motivation of this work. We see that this term breaks the equivariance for some values. First and foremost, we see that the symmetry breaking term has an exponentially vanishing amplitude. Then, this term becomes even smaller when $f_x \simeq f_y$, which is, in fact, the case on hardware. These two results combined indicate that the noise-induced symmetry breaking should not hinder the equivariance under Pauli channels. Last but not least, the value of $\theta$ also plays a role in the amount of symmetry breaking. It may, in fact, make the symmetry breaking zero regardless of the values of $f_x$ and $f_y$. This is a natural result, as there will be decompositions which improve robustness against noise. However, such decompositions of gates may not be available on hardware, and one should keep this in mind during the transpilation process. 

Now, let's consider the non-unital AD channel with the probability $\gamma$. Then, the outcome of the circuit can be written as,

\begin{align}
\hat{y}(x) &= \frac{1}{4}((1-\gamma)^{d/2}+(1-\gamma)^{d})\cos(x) \nonumber \\
        & + \frac{1}{4}((1-\gamma)^{d/2}-(1-\gamma)^{d})\cos(x+2\theta) \nonumber\\
        & + \frac{1}{2}\Lambda_{\mbox{\tiny AD}_{(4,1)}}^{(d)} \sin(\theta).
\label{eq:toy-ad}
\end{align}

The term $\Lambda_{\mbox{\tiny AD}_{(4,1)}}$ refers to the only off-diagonal entry in the Pauli Transfer Matrix (PTM) of the AD channel. The upper index $(d)$ denotes the $d^{th}$ power of this matrix. We refer the reader to Appendix~\ref{app:ptm} and \ref{app:toy} for the details. We can write this term explicitly as,

\begin{equation}
    \Lambda_{\mbox{\tiny AD}_{(4,1)}}^{(d)} \simeq d\gamma - \frac{d(d-1)}{2}\gamma^2.
\end{equation}

Here, we skip writing the remaining terms as their contribution will be negligible as long as we consider shallow circuits. Going back to the full expression for $\hat{y}(x)$, we immediately see that the AD channel results in a more complicated form. Nonetheless, it is easy to see the implications of each term one by one and this will give us the necessary intuition for the remaining part of this work. 

The first and second terms jointly result in the exponential concentration induced by the AD channel. This can be easily seen by setting $\theta=0$. The concentration happens around the third term, which shifts with the addition of each layer. This shift behaves approximately linear for practically relevant depths and noise levels\footnote{Current superconducting hardware has $\gamma \simeq 10^{-2}$ and CNOT depth of $10-20$. The values are approximate and vary from device to device.}, \textit{e.g.} $\mathcal{O}(\gamma d)$. The second term is the one responsible for symmetry breaking. The term $((1-\gamma)^{d/2}-(1-\gamma)^{d})$ behaves similar to the $\Lambda_{\mbox{\tiny AD}_{(4,1)}}^{(d)}$ term, \textit{e.g.} is approximately linear for relevant values of the parameters. Furthermore, it is upper bounded by $\mathcal{O}(\gamma d)$, and thus, the symmetry breaks approximately linearly in the number of layers $d$ or noise strength $\gamma$ under the AD channel.

One final important setting to consider is the combination of the Pauli channel with the AD channel. It is straightforward to compose this effective channel using the PTM picture. We obtain the noisy prediction as,

\begin{align}
\hat{y}(x) &= \frac{1}{4}(f_x^{d}(1-\gamma)^{d/2}+f_z^{d}(1-\gamma)^{d})\cos(x) \nonumber \\
        & + \frac{1}{4}(f_x^{d}(1-\gamma)^{d/2}-f_z^{d}(1-\gamma)^{d})\cos(x+2\theta) \nonumber\\
        & + \frac{1}{2}\Lambda_{\mbox{\tiny P+AD}_{(4,1)}}^{(d)}\sin(\theta),
\label{eq:toy-p+ad}
\end{align}

and the term $\Lambda_{\mbox{\tiny P+AD}_{(4,1)}}^{(d)}$ reads,

\begin{equation}
    \Lambda_{\mbox{\tiny P+AD}}^{(d)} \simeq \left(  \sum_{k=1}^d f_z^{k}  \right)\gamma - \left(  \sum_{k=1}^d (k-1) \times f_z^{k}  \right)\gamma^2.
\end{equation}

This term determines the shift of the mean. We see that it behaves the same except it is this time modulated with the Pauli fidelity $f_z$ at every layer. Similarly, the amplitude of symmetry breaking depends on the second term as follows,

\begin{align}
\hat{y}(x) &- \hat{y}(-x) =  \nonumber \\
& -(f_x^{d}(1-\gamma)^{d/2}-f_z^{d}(1-\gamma)^{d})\sin(\theta)\sin(x)/2
\label{eq:toy-p+ad-symm}
\end{align}

This means the symmetry breaking is also modulated with the Pauli fidelity $f_x$ and $f_z$ in each layer. Notice that we can recover the term for pure AD channel if we set $f_x=f_z=1$. Overall, the behavior of the term doesn't change, and it grows approximately linear in AD channel noise strength $\gamma$ with minor contributions from the Pauli channel. This statement can also be generalized to multi-qubit systems. Following the structure of Eq.~\ref{eq:2q-bitflip}, we see that the addition of local noise channels on other qubits has negligible effects as these terms appear as multiplicative terms. Hence, we conjecture that a generic EQNN model experiences symmetry breaking dominantly under the AD channel, and the amount grows linearly in noise strength $\gamma$ and depth $d$. 

In Section~\ref{sec:experiments}, we perform numerical experiments to confirm the implications of the toy model and present evidence directly from hardware runs. For this purpose, we continue by introducing metrics that can be computed using the simulation and hardware data such that we can decouple the symmetry breaking terms from the rest of the terms in the model outputs. 

\subsection{Quantifying symmetry breaking}
\label{sec:lm}

Preserving symmetries and quantifying the amount of symmetry are paramount for the success of tasks such as state preparation and time evolution of quantum systems in the presence of hardware noise. In fact, there is a growing literature that studies these aspects~\cite{tran_faster_2021, nguyen_digital_2022}. Although this may look like a very similar problem in GQML, there is a fundamental difference. In the former, the state belongs to a subspace that is governed by the symmetry of the corresponding system, while in the latter, what matters is the relative positions of the symmetric inputs in the subspace that is governed by the label symmetry. Furthermore, in tasks such as binary classification, the continuous output of a model is mapped to a binary decision based on a threshold. This means that small deviations in the expectation value may not change the binary decision. Overall, these points relax the conditions to preserve the symmetry in the context of GQML. Ragone et al.~\cite{ragone2023unified} recently introduced \textit{$\mathfrak{g}$-purity}, which can be used to measure the symmetry breaking in GQML, but $\mathfrak{g}$-purity is expensive to compute and doesn't account for the binary decisions. Thus, there is a need to define metrics that can capture all of these aspects.

We start by defining a metric that can use the continuous outputs of a model (\textit{i.e.} $\hat{y}_i$ for input $\mathbf{x}_i$\footnote{Bold symbols are used to represent vectors. Here $\mathbf{x}_i$ denotes $i^{th}$ data sample with arbitrary size.}). For this purpose, we have to make a choice of the symmetry group. In this paper, we focus on the discrete $\mathcal{Z}_2=\{e,\sigma\}$ symmetry, such that $R(e)\cdot(\mathbf{x}_i)=(\mathbf{x}_i)$ and $R(\sigma)\cdot(\mathbf{x}_i)=(\mathbf{x}_j)$, where $R$ is the representation of the symmetry group element in the data space $\mathcal{R}$. Then, the equivariance implies $\hat{y}_i = \hat{y}_j$. We define accordingly $\mathcal{Z}_2$ symmetry generalized McNemar-Bowker (MB) test~\cite{krampe2007bowker} as follows,

\begin{definition}[$\mathcal{Z}_2$ generalized MB test] Consider the $\mathcal{Z}_2=\{e,\sigma\}$ symmetry, such that $R(e)\cdot(\mathbf{x}_i)=(\mathbf{x}_i)$ and $R(\sigma)\cdot(\mathbf{x}_i)=(\mathbf{x}_j)$. Then, the normalized McNemar-Bowker (MB) test~\cite{krampe2007bowker} of a model with predictions $\hat{y}_i$ for input $\mathbf{x}_i$ over $M$ samples can be defined as,
\begin{equation}
    \chi^2 = \frac{1}{M}\sum_{i=1}^M \frac{(\hat{y}_i-\hat{y}_j)^2}{\hat{y}_i+\hat{y}_j}
\end{equation}
\end{definition}

This definition can be further extended to the binary predictions. For this purpose, we define the \textit{threshold function} $\tau$, which is a step function that has the transition point $t$. A na\"ive choice for the value of $t$ is the center point of the two binary class predictions (\textit{e.g.} $t=0.5$ if the classes are defined as 0 and 1, $t=0$ if the classes are defined as -1 and 1). However, as we illustrated earlier, the predictions of a model may shift towards a value under hardware noise, and thus, the central and fixed $t$ value becomes a bad choice. Furthermore, this value is often optimized by following the area under the curve of the receiver operation characteristics of a model~\cite{bradley1997use}. Unsuitably, this makes the choice data-dependent. With these points in mind, we choose the threshold $t$ such that it is the median of the continuous outputs of a model for the inputs from the training set. This allows us to update the value and account for the shift in the center of the expectation values. Then, we can use the binary predictions $\tau(\hat{y}_i)$ to compute $\chi^2$. We will refer to this value as \textit{label misassignment} (LM), as it counts the amount of the predictions that have a different prediction than their $\mathcal{Z}_2$ counterparts. 

\begin{definition}[Label Misassignment (LM)] Consider the $\mathcal{Z}_2=\{e,\sigma\}$ symmetry, such that $R(e)\cdot(\mathbf{x}_i)=(\mathbf{x}_i)$ and $R(\sigma)\cdot(\mathbf{x}_i)=(\mathbf{x}_j)$. Let us take a  model returning binary predictions $\tau(\hat{y}_i)$, where $\hat{y}_i$ are the continuous predictions of the model for input $\mathbf{x}_i$ and $\tau$ a step function that has the transition point at the median of all $\hat{y}_i$. Then, label misassignment (LM) of a model over $M$ samples can be defined as,

\begin{equation}
    \mbox{LM} = \frac{1}{M}\sum_{i=1}^M \frac{(\tau(\hat{y}_i)-\tau(\hat{y}_j))^2}{\tau(\hat{y}_i)+\tau(\hat{y}_j)}
\end{equation}
\end{definition}

Notice that each term in the sum is either 0\footnote{We avoid division by zero in the case of zero predictions, by adding a small epsilon to the denominator for the numerical experiments} (if the model prediction is the same for $\mathbf{x}_i$ and $\mathbf{x}_j$) or 1 (if the predictions are different). This allows LM to count the amount of misassigned predictions. For example, a model that has perfectly symmetric outputs will be 0\% of LM, while a model that produces random outputs 50\% of LM. A model that predicts the opposite label for all symmetric inputs will have 100\% of LM. 

Furthermore, $1-\mbox{LM}/2$ can be used to upper bound the accuracy of a model. Consider the model that predicts the opposite label each time (\textit{i.e.} LM=1.0); this model can have, at best, 50\% accuracy. Similarly, a model with random outputs (\textit{i.e.} LM=0.5) can't have an accuracy larger than 75\%. Notice that $1-\mbox{LM}/2$ doesn't predict the accuracy of a model but only upper bounds it, otherwise one would expect the completely random model to have 50\% accuracy.

\section{Experiments}
\label{sec:experiments}

In this section, we provide numerical experiments to validate our findings. To achieve this goal, we perform binary classification experiments, compute $\chi^2$ and label misassignment (LM) that we previously defined in Section~\ref{sec:lm}, utilizing both simulated and hardware results. 

\begin{figure}[!h]
    \includegraphics[width=.7\linewidth]{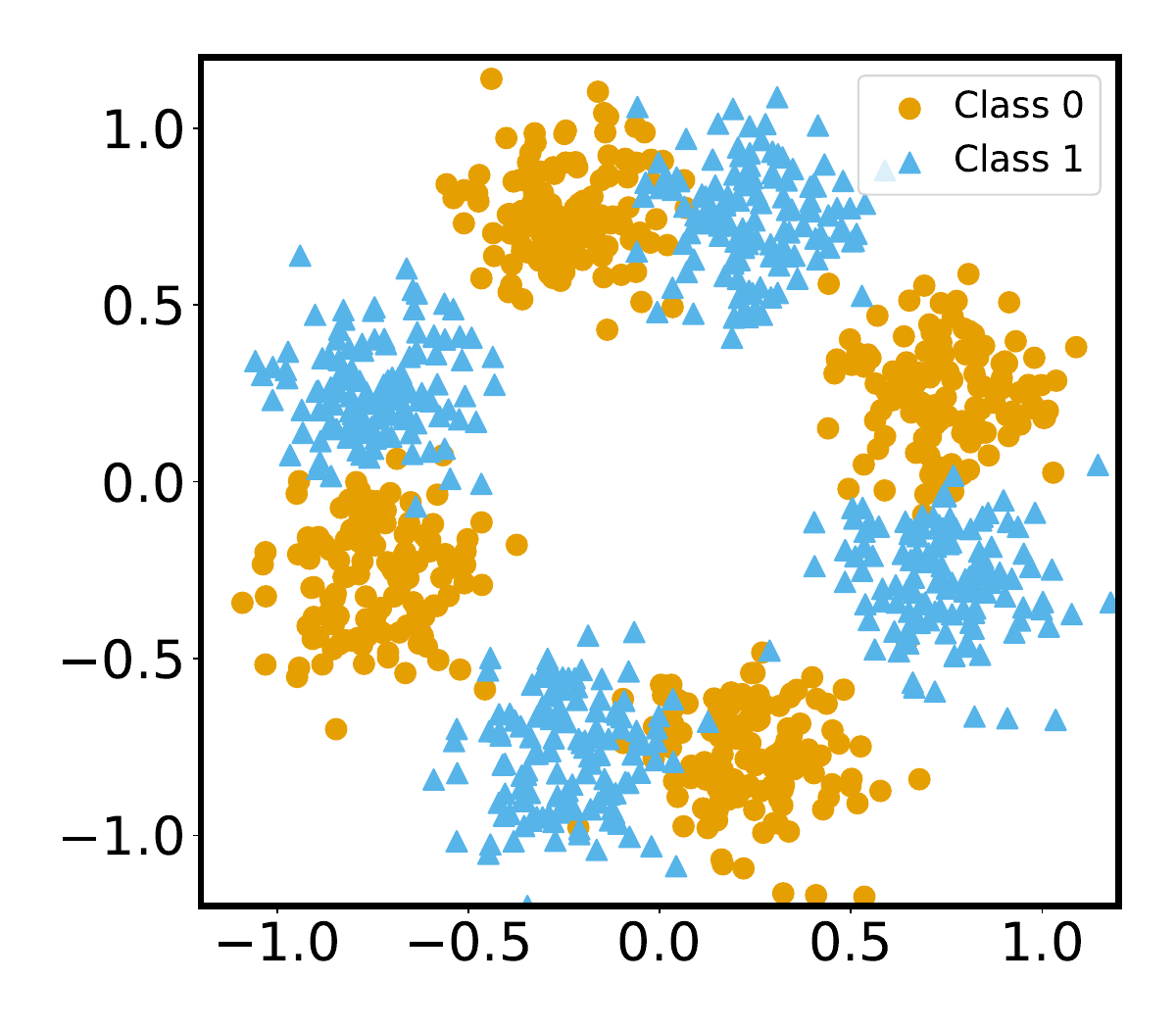}
    \caption{An ad-hoc dataset with $\mathcal{Z}_2$ label symmetry such that $R(\sigma)\cdot(\mathbf{x}_i)=-\mathbf{x}_i$.}
    \label{fig:2q-dataset}
\end{figure}

For the experiments, we consider datasets with $\Z$ symmetry as described before. Accordingly, we choose the symmetry transformation such that $R(\sigma)\cdot(\mathbf{x}_i)=-\mathbf{x}_i$. We generate a dataset, as depicted in Fig.~\ref{fig:2q-dataset} that carries this symmetry for the classification experiments. 

\begin{figure}[!t]
	\centering
\subfloat[\label{fig:eqnn-z} EQNN-Z]{\includegraphics[width=\linewidth]{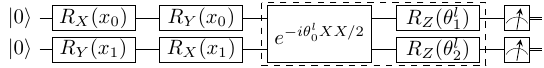}} 
\newline 
\subfloat[\label{fig:eqnn-xy} EQNN-XY]{\includegraphics[width=\linewidth]{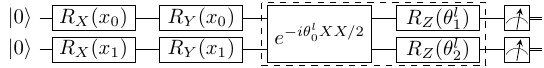}} 
\newline 
\hspace*{\fill}
\subfloat[\label{fig:xx} XX decomposition]{\includegraphics[width=0.85\linewidth]
{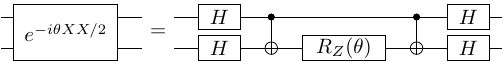}}
\hspace*{\fill}
\newline 
\subfloat[\label{fig:bel} BEL]{\includegraphics[width=0.85\linewidth]{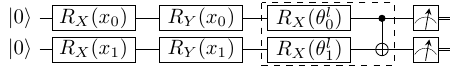}}
	\caption{Two qubit circuits used in the experiments.}
	\label{fig:2q-circuits}
 \end{figure}

As we illustrated earlier, the choice of an equivariant data embedding induces a specific unitary representation of the symmetry group element, which will restrict the choices of the parametrized gates and the observable. We define two different two-qubit EQNN models, \textit{EQNN-Z} and \textit{EQNN-XY}, as shown in Fig.~\ref{fig:eqnn-z},~\ref{fig:eqnn-xy}. In both models the data encoding is performed with the Pauli rotation gates $R_Y$ and $R_X$, inducing the representation $R_q(\sigma) = Z_0 Z_1$. EQNN-XY data encoding uses the same gate at each layer, while in the EQNN-Z case, the order of $R_X$ and $R_Y$ gates are alternated. 

The parametrized gates in both cases are the same. We select three generators $G \in \{ X_0 X_1, Z_0 I_1, I_0 Z_1 \}$ from the set of commutators of the representation $Z_0 Z_1$. These generators are used to obtain the parametrized gates of the form $U_G = \mbox{exp}(-i\theta G/2)$. It is sufficient to use only these three generators, because the nested set of commutators of these three generators is equivalent to the set of commutators of the representation $Z_0 Z_1$. The three gates form a parametrized layer and each layer is repeated $d$ times, having independent parameters. Lastly, we choose the equivariant observable $O = (Z_0+Z_1)/2$ for EQNN-Z ansatz and $O = X_0 Y_1$ for EQNN-XY. 

Spurious symmetries may arise when building EQNN models. This appears as an unwanted $SWAP$ symmetry (\textit{i.e.} $x_i^0 \rightarrow x_i^1,~x_i^1 \rightarrow x_i^0$) in our example. We handle this in different ways in two models. EQNN-XY breaks the unwanted symmetry by employing the observable $X_0 Y_1$, which doesn't commute with the $SWAP$ gate. In the case of EQNN-Z, this is broken at the data encoding level, since the order of $R_X$ and $R_Y$ gates are alternated. The choice of breaking it at the data encoding level or the measurement level will impact the performance of the model, as we will see later.

Additionally, we define a non-equivariant model that doesn't use any geometric priors from the dataset as shown in Fig.~\ref{fig:bel}. We denote this model with \textit{BEL} and compare it to the EQNN models using the same observables. 

Last but not least, to model the effect of noise for EQNN circuits, we decompose the $\mbox{exp}(-\theta XX/2)$ gate using a CNOT based decomposition as depicted in Fig.~\ref{fig:xx} and apply noisy gates after each layer as it was shown in Fig.~\ref{fig:noise-model}. Furthermore, we simulate the EQNN-Z circuit without any decomposition to discern the noise effect and refer to this experiment as \textit{EQNN-Z-native}.

\subsection{Binary classification}
\label{sec:2q-train}

\begin{figure*}[!t]
	\centering
\subfloat{\label{fig:DP_accuracy}\includegraphics[width=0.33\linewidth]{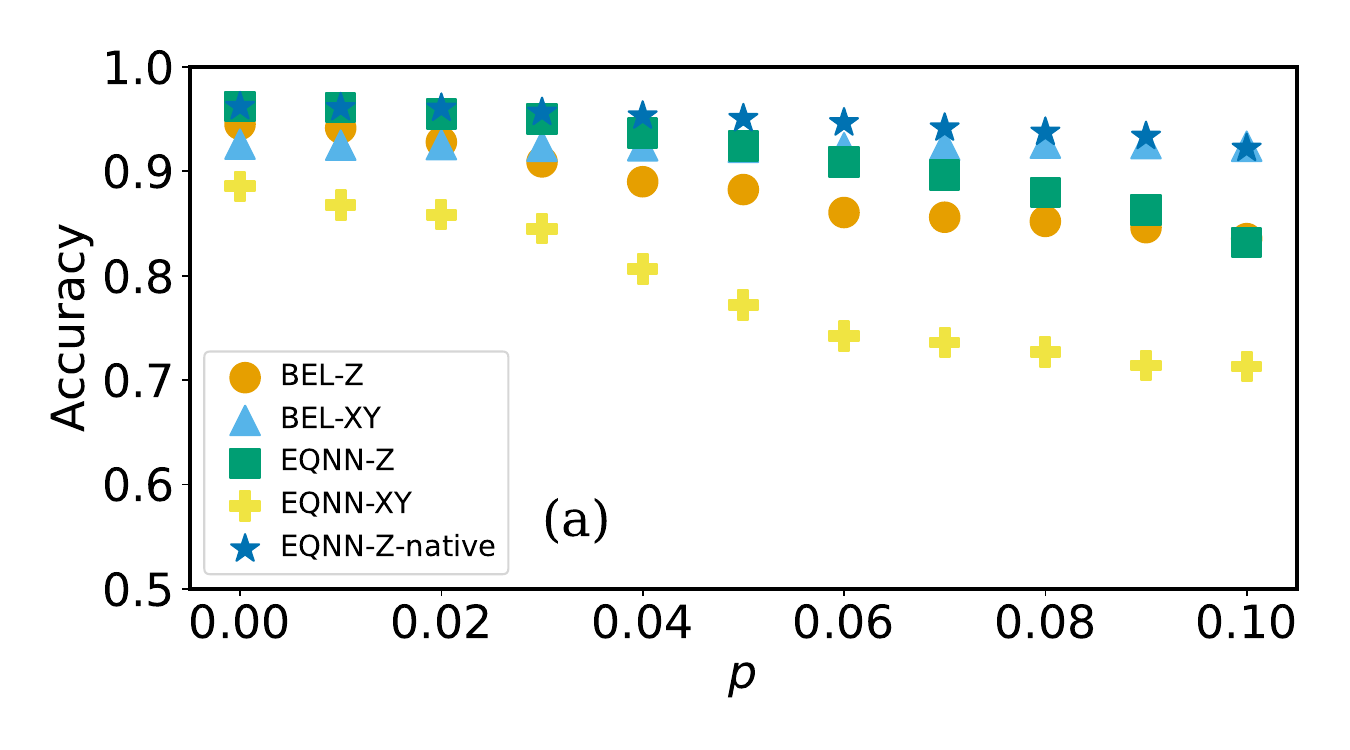}} 
\subfloat{\label{fig:AD_accuracy}\includegraphics[width=0.33\linewidth]{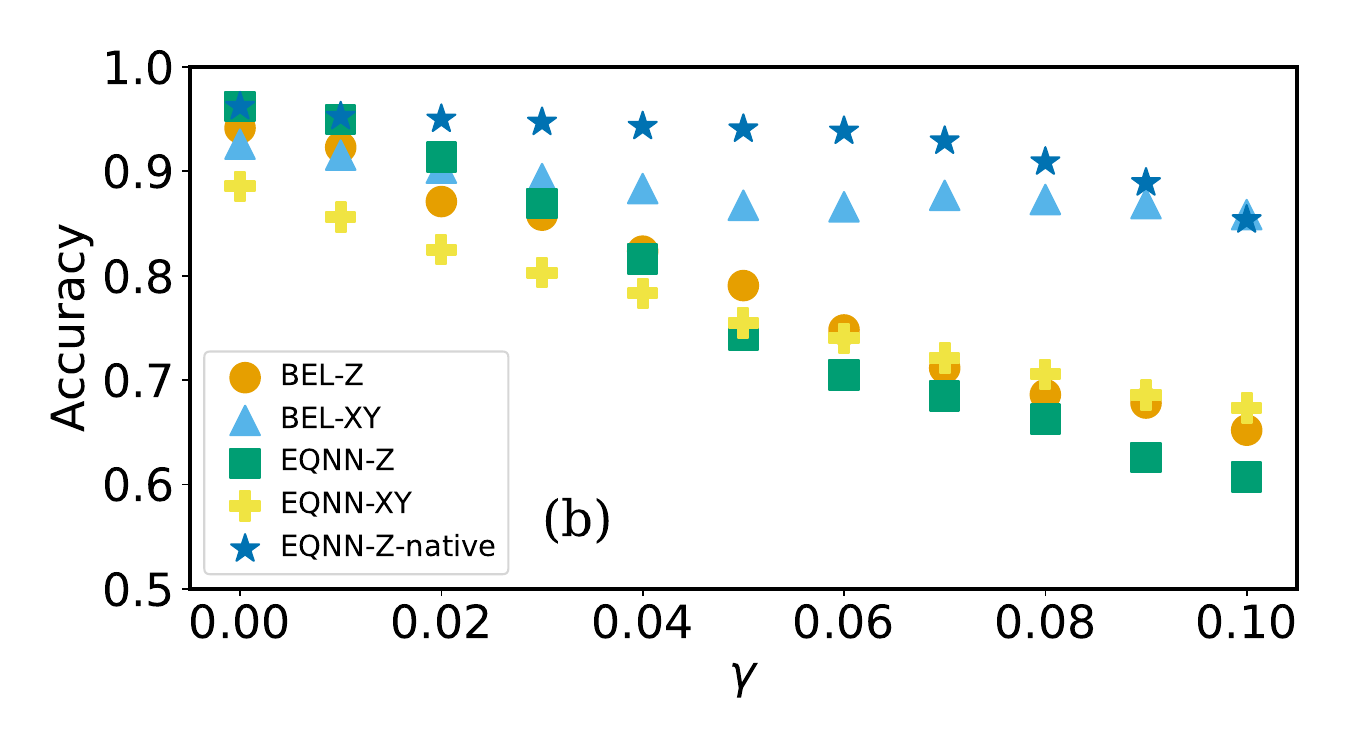}} 
\subfloat{\label{fig:AD_with_AT_accuracy}\includegraphics[width=0.33\linewidth]{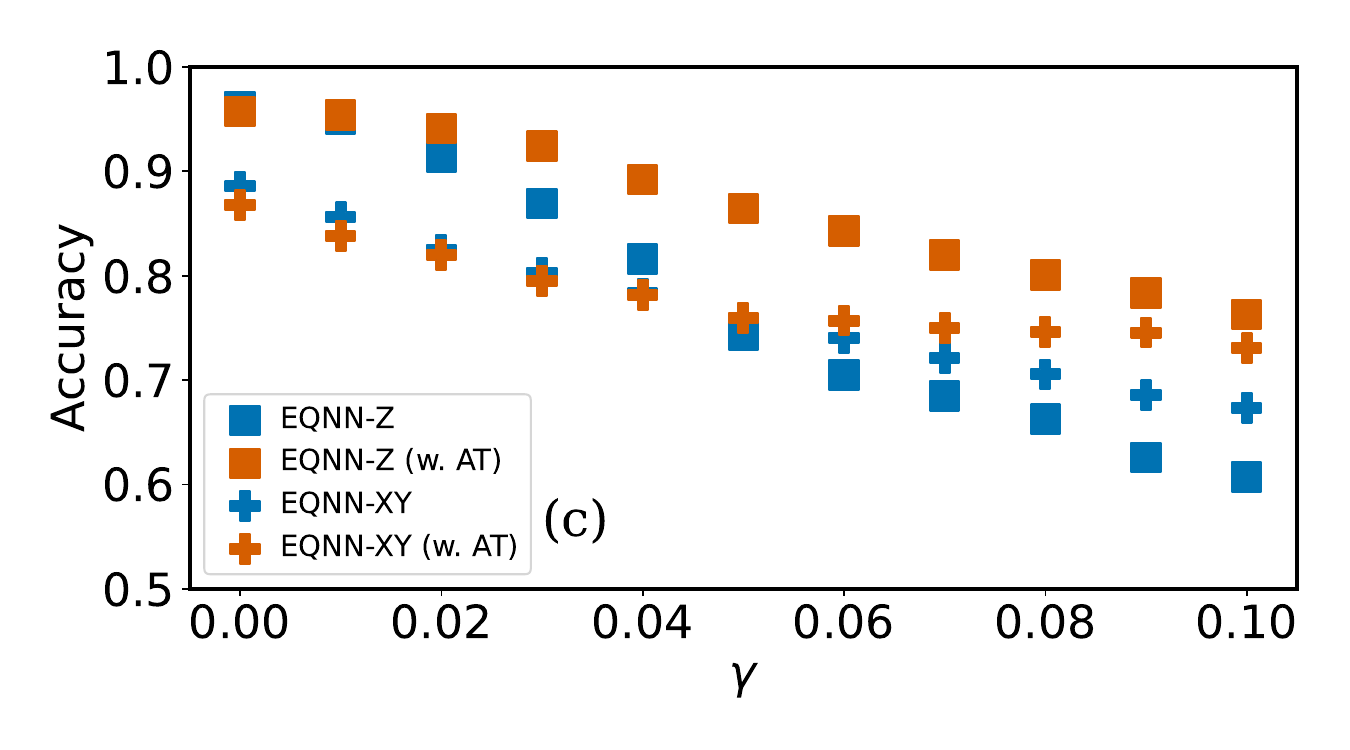}} 
	\caption{Binary classification results under noise channels. All models are trained with ten different initializations and layers varied from 1-10. The test accuracy, averaged over the runs, is plotted for the best-performing layer of the corresponding model. Noise strength $p$ in the case of the DP channel and $\gamma$ in the case of the AD channel is varied from 0.0 to 0.1 with 0.01 increments. a) Results under DP channel. b) Results under AD channel. c) Results under AD channel with and without using adaptive thresholding (AT) during training.}
	\label{fig:2q-training}
\end{figure*}

Numerical experiments for classification are conducted using two qubit circuits, described previously. To compare the accuracy of the model under different noise channels, we run all circuits up to ten layers for a given noise strength and plot the value of the best-performing layer averaged over ten runs. This is to find the best-case scenario for each model as each model will have different effective depth and experience noise differently for a given number of layers $d$. The binary cross entropy loss function was minimized using the Adam optimizer~\cite{Kingma2014AdamAM}. The simulations are performed in the absence of shot noise using the Python library Pennylane~\cite{bergholm2018pennylane}.

We showcase the results of trained models under varying strength of DP and AD channels respectively in Fig.~\ref{fig:DP_accuracy},~\ref{fig:AD_accuracy}. In the absence of noise, all models can show more than 90\% accuracy. We see a discrepancy between the EQNN-XY and EQNN-Z models. This is due to the location of the spurious symmetry breaking we mentioned earlier. Since the EQNN-Z model breaks this spurious symmetry at the data encoding level, it is more expressive and, hence, can perform better.

In the case of the DP channel, all models experience a similar performance drop. This is a natural outcome of the gradients getting smaller as noise strength increases due to the emergence of noise-induced barren plateaus. 

When considering the AD channel, the performance drops more significantly, characterized by a sharper decline in accuracy. In particular, the BEL-Z, EQNN-Z, and EQNN-XY demonstrate more pronounced effects compared to other models, while BEL-XY performs the best among the four models. There are two reasons for this. The first reason is the symmetry breaking, which impacts both EQNN models. This effect can be observed better when we compare EQNN-Z-native and EQNN-Z. Our intuition from the toy model was that the symmetry breaking should be observed in the case of the AD channel and not in the DP channel. We observe that under the DP channel, these models perform much similarly than they do under the AD channel. Since EQNN-Z-native results in shorter depth, it is expected to perform better also under DP channel.

The second reason is the shift of mean for the $Z$ observable under the AD channel. This results in the model having a bias towards one label, when a fixed threshold function is used. To alleviate the effects of the shift of mean, a simple practical trick called adaptive thresholding is employed. Using prior knowledge on the dataset labels (\textit{e.g.} a balanced dataset has equal amounts of both classes), one can adaptively change the prediction threshold throughout training. The threshold value can be computed as the median over the predictions of the training set at every iteration. Our results depicted in Fig.~\ref{fig:AD_with_AT_accuracy}, indicate significant improvement in the model performance, particularly in the case when measurements are affected asymmetrically in $z$-direction. Consequently, this improvement would not be limited to equivariant models. This result shows that adaptive thresholding is a useful and cheap technique to improve model performance for binary classification under hardware noise.

One final point worth noting is the exceptional performance of the EQNN-Z-native model. It consistently outperforms all other models under both the DP and AD channels. The impact of the DP channel on both equivariant and non-equivariant models is expected to be similar. However, what stands out is that the EQNN-Z-native model shows no significant performance drop under the AD channel. This resilience is attributed to the specific choice of the $Z_0 Z_1$ representation, which commutes with the AD channel (please see Appendix~\ref{app:noise} for details). Despite its impressive performance, it's important to note that this model faces implementation challenges on current quantum hardware due to limitations in the native gate set.

\subsection{Symmetry breaking}
\subsubsection{Two qubit case}
\label{sec:2q-symm}

In order to explain the discrepancy of performance in training, we measure the proposed metrics $\chi^2$ and LM using simulated data as well as data collected from superconducting quantum computers.

\begin{figure}[!t]
    \centering
    \includegraphics[width=\linewidth]{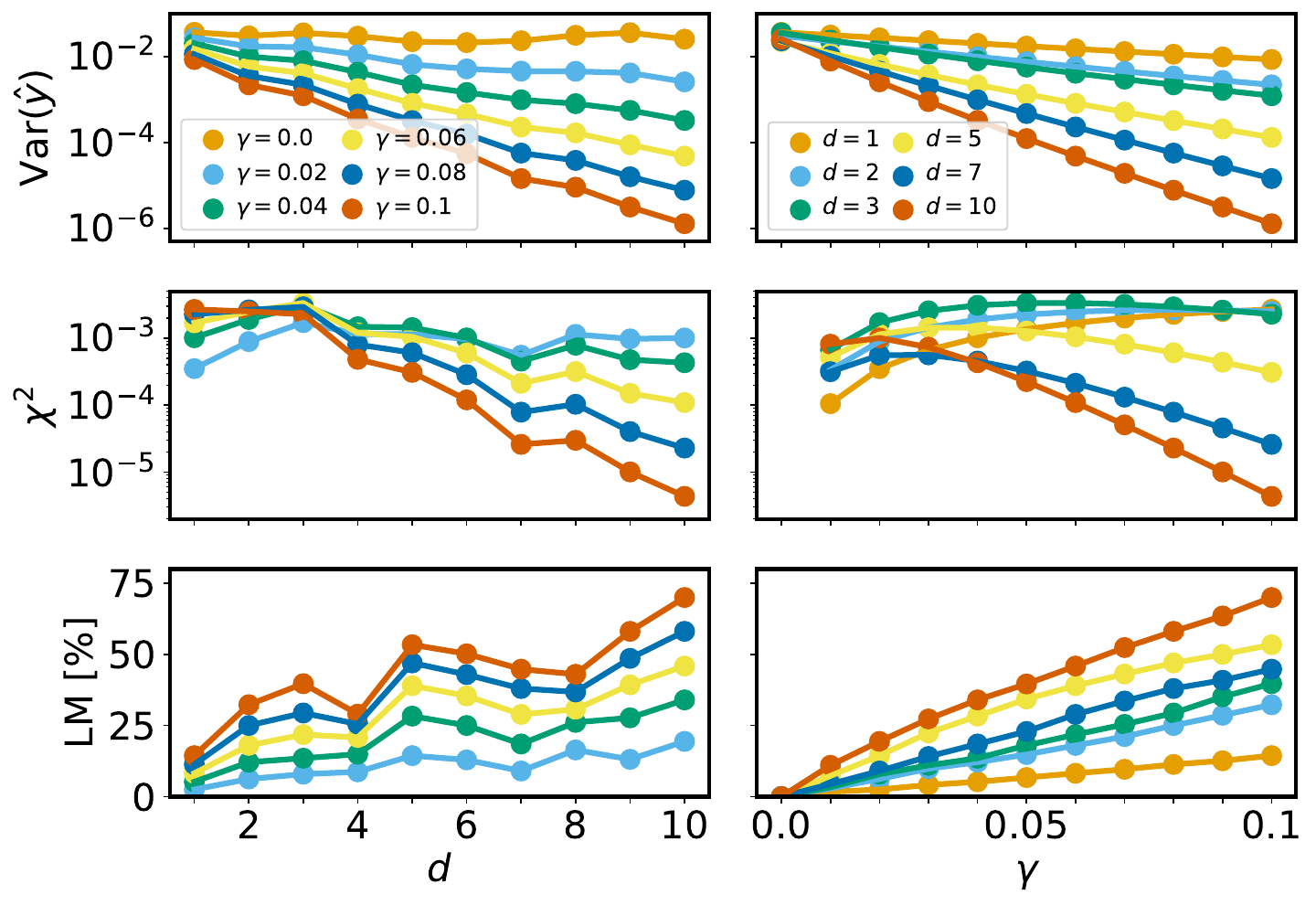}
    \caption{Simulated two-qubit symmetry breaking for the EQNN-XY model under AD channel. Both columns show the same data points, on the left metrics are plotted against number of layers $d$, on the right metrics are plotted against noise strength $\gamma$.}
    \label{fig:2q-symm-sim}
\end{figure}

We start by considering the two-qubit EQNN-XY model and collect predictions with ten random initializations for 400 input data samples under different strengths of the AD channel. We plot the variance of the output predictions, $\chi^2$, and LM averaged over the ten runs for varying number of layers in Fig.~\ref{fig:2q-symm-sim}. The exponential decay of the variance numerically confirms the existence of the noise-induced BPs. The value of $\chi^2$ first increases and then decreases for small values of $\gamma$ and completely decreases for larger values. This is a joint result of symmetry breaking and noise-induced BPs. $\chi^2$ can measure the symmetry breaking until the exponential concentration dominates the landscape and brings all predictions closer to the same value. In fact, it is upper-bounded by the variance. One can use the LM metric to decouple these two effects. LM can measure the symmetry breaking separately since it uses the adaptive threshold $t$. LM grows linearly in the noise strength $\gamma$ and the number of layers $d$. This perfectly matches the analytical expression we have obtained in Eq.~\ref{eq:toy-ad} and gives numerical evidence for our linear symmetry breaking conjecture.

\begin{figure}[!t]
    \centering
    \includegraphics[width=\linewidth]{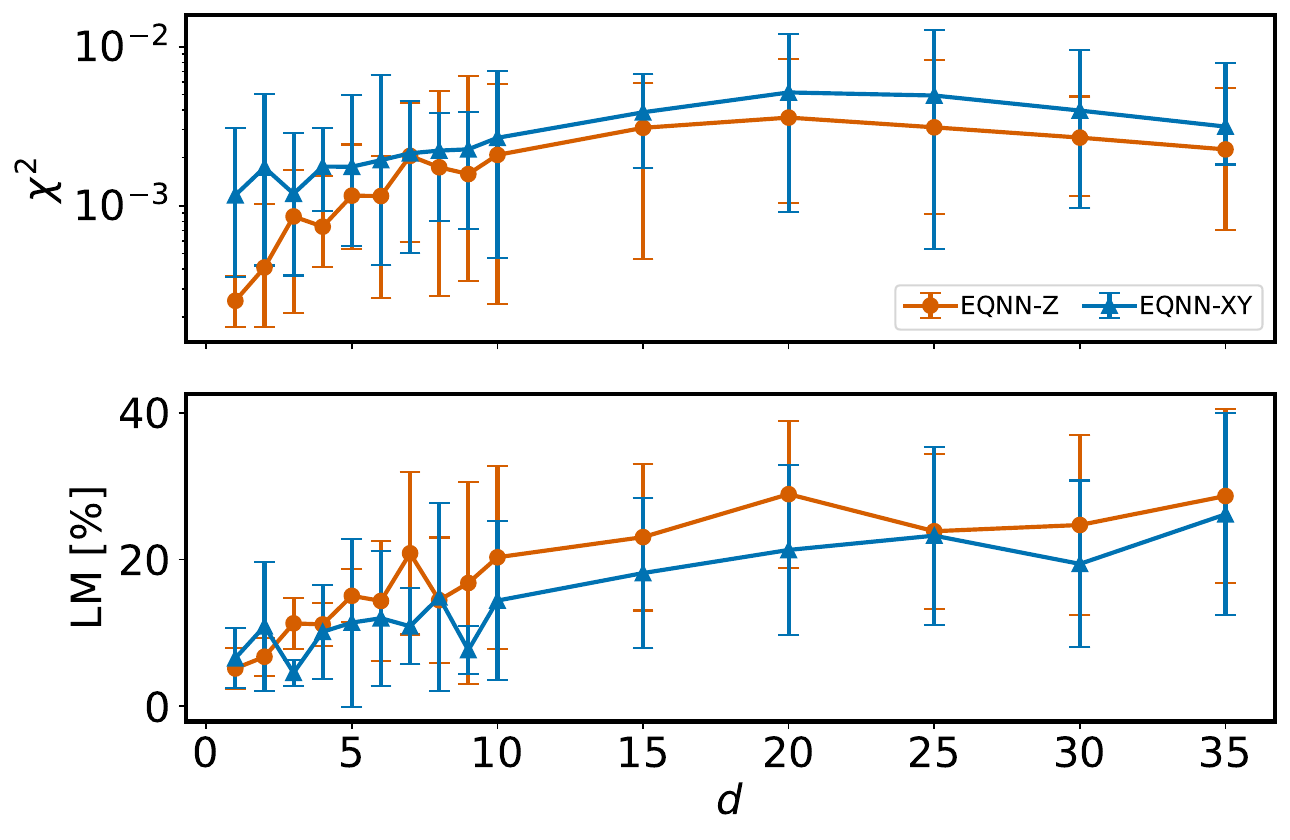}%
    \caption{Two qubit symmetry breaking for the EQNN-XY and EQNN-Z models measured on the \cairo~superconducting quantum computer. (top) $\chi^2$ and (bottom) LM are plotted against the number of layers $d$.}
    \label{fig:2q-symm-hw}
\end{figure}

Using this result, we can also comment on the binary classification performance. Following the bottom right panel of Fig.~\ref{fig:2q-symm-sim}, we see that LM reaches 20\% in the shortest depth scenario. We can use this line to compare the performance of the EQNN-XY model. As mentioned earlier, LM upper bounds the accuracy with $1-\mbox{LM}/2$. Based on this, we can say that at $\gamma=0.1$, the EQNN-XY model should experience a 10\% drop in accuracy, only caused by symmetry breaking. It is difficult to comment on the impact of a single factor, as there are many factors that contribute to the drop in performance in the presence of noise. Nonetheless, looking at Fig.~\ref{fig:2q-training}, this value appears reasonable.

Next, we repeat this experiment on the \cairo~superconducting quantum computer using the models EQNN-Z and EQNN-XY with 4000 shots. For this purpose, we use the same dataset and the same parameters for the circuits. We report $\chi^2$ and LM values for the number of layers up to 35 in Fig.~\ref{fig:2q-symm-hw}. These results show that both models behave similarly, matching the numerical simulations that were conducted only using the AD channel. This confirms our prediction of the fact that the AD channel dominantly contributes to the symmetry breaking for this setting. 

There is a discrepancy between the $\chi^2$ and LM values of the two models. In the case of $\chi^2$, both models observe the increase and then later the decrease due to concentration. However, it's not enough to look at the value of $\chi^2$ to make comments on the amount of symmetry breaking. This is because the scale of this metric is controlled by the variance of the observable, and one should keep this in mind when comparing observables with different variances. Next, looking at the LM plot, we see that the EQNN-Z model, in general, suffers more symmetry breaking compared to the EQNN-XY model. This is mainly due to the fact that the $z$-direction being asymmetric in the AD channel. This result also agrees with Fig.~\ref{fig:2q-symm-hw}, in which EQNN-Z performance detoriates faster. Furthermore, we observe that LM behaves linearly in the number of layers while approaching 50\%. The LM values this time converge to 50\% since we have shot noise, and the output becomes completely random at a large depth. All of these results combined align well with the predictions of the AD channel dominating the symmetry breaking.

\subsubsection{Multi-qubit case}
\label{sec:Nq-symm}

So far, we have considered only the two-qubit case in our experiments, yet our primary interest revolves around the behavior of symmetry breaking at a large scale. Performing simulations on a larger scale imposes significant challenges, becoming computationally expensive. In this section, we focus on obtaining empirical results from the 127-qubit \cusco~superconducting chip. For this purpose, we use the nearest neighbor qubits as shown in Appendix~\ref{sec:top}. 

\begin{figure}[!h]
	\centering
	\includegraphics[width=\linewidth]{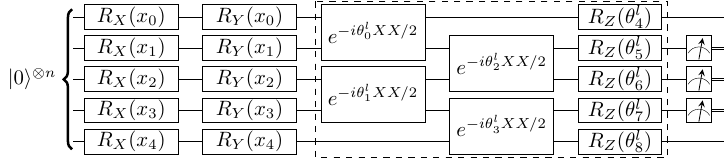}
	\caption{Hardware efficient circuit used in the experiments to measure symmetry breaking. The part inside the dashed box is repeated $d$ times with different parameters. Here, a five qubit circuit is plotted for reference. This model is denoted with EQNN-HWE.}
	\label{fig:Nq-circuits}
\end{figure}

\begin{figure}[!t]
    \centering
    \includegraphics[width=\linewidth]{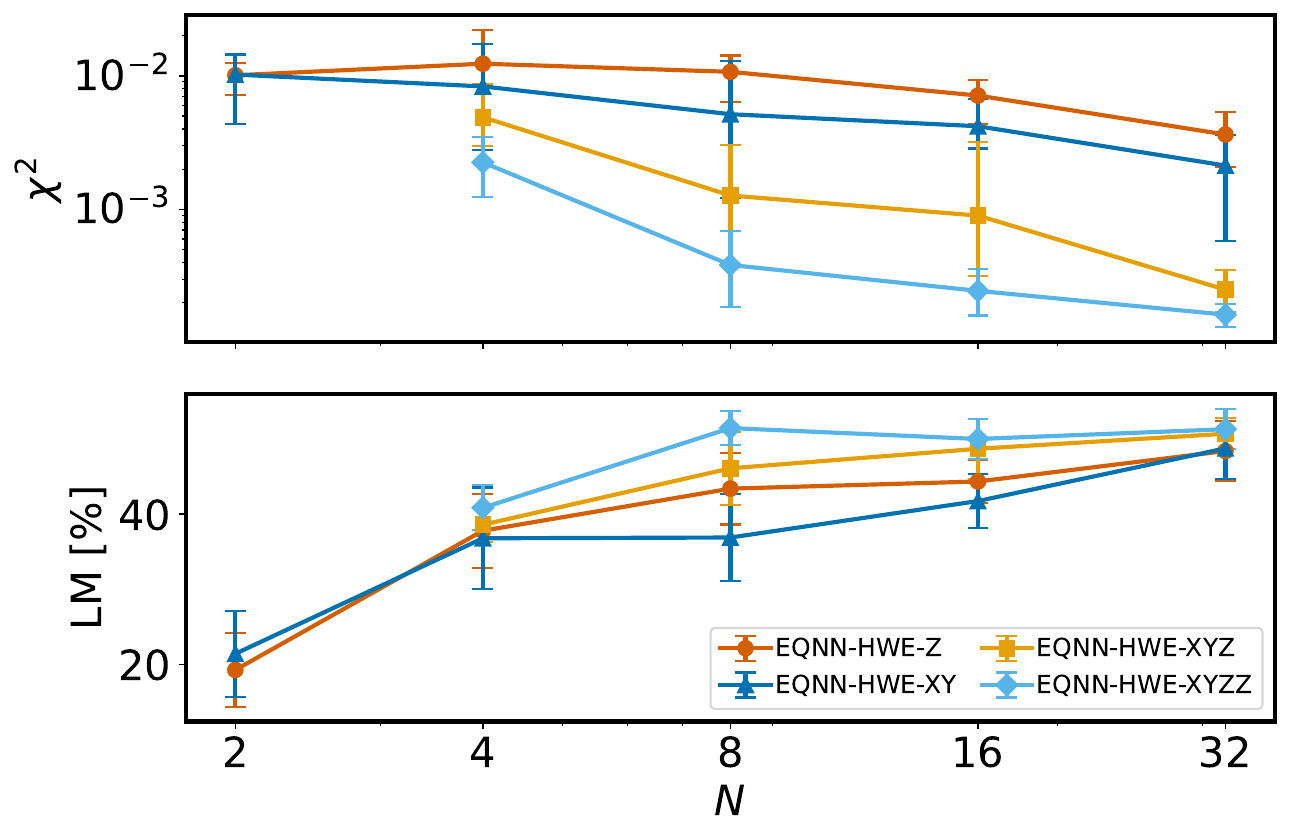}%
    \caption{Log-depth EQNN-HWE results from \cusco. Hardware efficient circuits defined in Fig.~\ref{fig:Nq-circuits} with the number of layers $d = \mbox{log}_2 n$. Each model uses a different observable denoted in the legend. The $x$-axis is plotted in log-scale, such that it is linear in number of layers.}
    \label{fig:Nq-symm-hw}
\end{figure}

In order to run experiments on hardware, we define a hardware efficient multi-qubit circuit, \textit{EQNN-HWE}, illustrated in Fig.~\ref{fig:Nq-circuits}. Data encoding is performed using $R_X$ and $R_Y$ gates. This results in the representation $Z^{\otimes n}$, similar to all other ans\"atze we studied so far. A hardware efficient brick-work layer, constructed from $\mbox{exp}(-\theta XX/2)$ gates, followed by $R_Z$ gates, is repeated $d$ times. Notably, observables are measured on central qubits to maximize the amount of gates captured by the light cone. Our experiments include probing observables with varying bodyness: $\{Z, XY, XYZ, XYZZ \}$. Notice that all the observables commute with the representation $Z^{\otimes n}$ to ensure equivariance. Fig.~\ref{fig:Nq-symm-hw} presents the $\chi^2$ and LM values obtained for log-depth ($d = \mbox{log}_2 n$) circuits with varying observables. 

Results obtained for $\chi^2$ highlight disparities in the bodyness of the observables. As the locality of the observable increases, the measured expectation values demonstrate a significantly accelerated concentration, leading to a decrease in $\chi^2$. This is expected as the locality of an observable is directly related to the variance of an observable~\cite{Cerezo2021CostBP, ragone2023unified} in general. We note that there are exceptions to this statement in the literature~\cite{diaz2023showcasing}. Furthermore, the trend for $\chi^2$ with respect to the number of qubits aligns with the two-qubit models that were simulated only using the AD channel. The behavior of LM is consistent with prior findings, showcasing that a log-depth equivariant circuit approaches almost 50\% in LM starting from $n = 8$ qubits, corresponding to random outcomes. 

These results indicate that log-depth EQNN models are not scalable on this hardware due to the combination of concentration and symmetry breaking. This shouldn't be surprising since there is always a cutoff depth for reasonable output on noisy devices. Although this cutoff depth does not look very promising, it can be further improved with various methods.

\textit{Pulse-efficient} implementation is one of the possible methods to improve the results at the hardware level. The default IBM Qiskit~\cite{Qiskit} transpilation only exposes fixed pulse gates, such as the calibrated CNOT gate, or \textit{ECR} gate, which is equivalent to CNOT gate up to single-qubit pre-rotations~\cite{Earnest2021, Egger2023PulseVAE}. Thus, any two-qubit gates are decomposed into a decomposition of CNOT and ECR gates and single-qubit gates. Although not ideal, this way of automated transpilation is less time-consuming and is a favorable application-agnostic approach. However, these fixed pulse gates have relatively long gate time for low entangling angles, and, thus leading to large errors. Thus, in order to improve the hardware result, it is possible to create $R_{ZX}(\theta)$ gates by controlling pulses in a continuous way, instead of using the fixed pulse gates. 

\begin{figure}[!t]
    \centering
    \includegraphics[width=\linewidth]{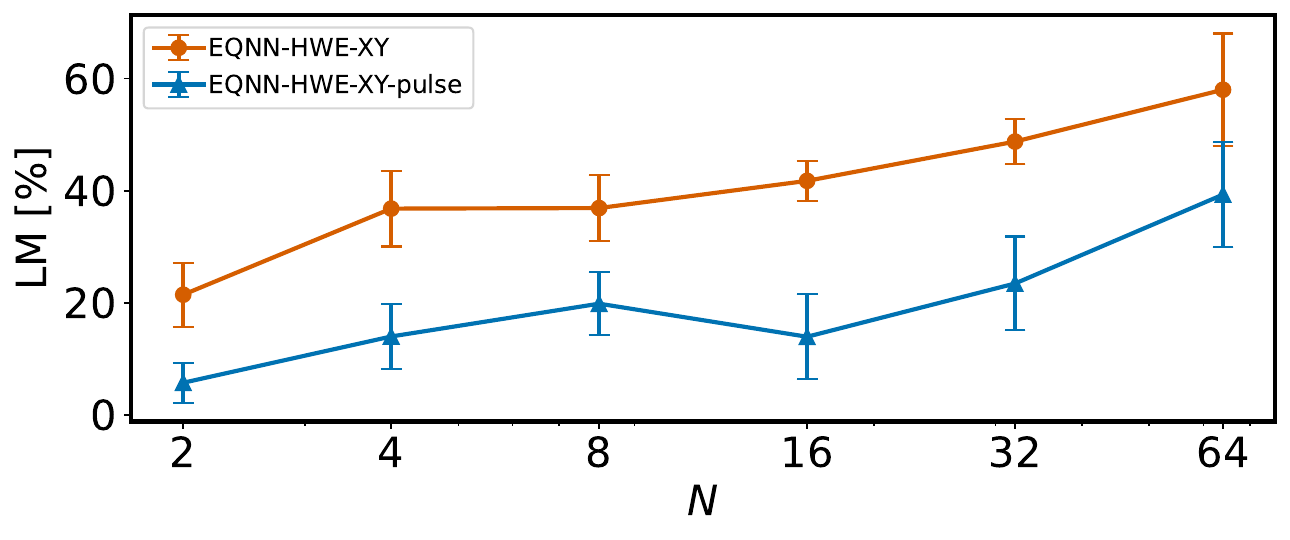}
    \caption{Label misassignment of the EQNN-HWE-XY model using different transpilation methods. The label EQNN-HWE-XY refers to the standard transpilation used throughout this work. The EQNN-HWE-XY-pulse refers to the pulse efficient transpilation~\cite{Earnest2021}.}
    \label{fig:pulse}
\end{figure}

Following Earnest et al.~\cite{Earnest2021}, we use the pulse-efficient implementation where the two-qubit quantum gates are decomposed into the hardware-native RZX gates. This allows us to implement the same circuit almost twice as fast using arbitrary parameterization of the pulse control. To show the effectiveness of this approach, we repeat the same experiment with EQNN-HWE-XY using this scheme and report the results in Fig.~\ref{fig:pulse}. As expected from our linearity argument, the symmetry breaking reduces to half of the previous experiment since twice faster execution can be thought as half the AD channel strength. We refer the reader to Appendix~\ref{sec:pulse} for more details on the pulse efficient execution.

\section{Conclusion}
\label{sec:conclusion}

In this work, we studied the behavior of EQNN models in the presence of noise. We highlight that these models experience symmetry breaking in the presence of realistic hardware noise. This adds another noise-induced complication to EQNN models, while the major one being noise-induced BPs~\cite{wang_noise-induced_2021}. Notably, we demonstrated that the impact of Pauli channels on symmetry breaking could be negligible, while the AD channel induces a symmetry breaking that is linear in the number of layers and noise strength. This further enables predicting the performance of an EQNN model on hardware prior to execution.

To address these challenges, we proposed effective strategies for mitigating performance drops caused by hardware noise. First of these is the adaptive thresholding that can cope with the concentration as well as the shift of mean. Furthermore, we showed that choosing the $Z^{\otimes n}$ representation is beneficial since it commutes with the AD channel. While our focus was on the $\Z$ symmetry for simplicity, our conclusions can be extended to other discrete symmetry groups. However, the implications for continuous groups remain uncertain and this makes it an interesting future research direction. Moreover, we demonstrated that more efficient hardware implementation can contribute to reducing symmetry breaking.

The symmetry protection under the Pauli channel result raises the question of employing Pauli twirling to convert non-unital noise channels to Pauli channels~\cite{van_den_berg_probabilistic_2023}. However, the scalability of the amount of twirls to preserve equivariance remains unclear, posing an open question for future exploration.

In our experiments, we haven't considered error mitigation methods. This was an intentional choice. Our target in this manuscript is to investigate the scalability of GQML on hardware, rather than just being able to execute circuits. This means error mitigation methods such as \textit{probabilistic error cancellation} (PEC) are not suitable for this study due to their exponential overhead~\cite{van_den_berg_probabilistic_2023}. Furthermore a na\"ive implementation of PEC may result in further loss of equivariance. This opens up new avenues to explore whether we can perform PEC by preserving given group symmetries. Additionally, we briefly explore the potential of \textit{zero noise extrapolation} (ZNE) in Appendix~\ref{app:zne}, revealing its effectiveness when provided with analytical expectation values but highlighting challenges with a limited number of shots.

In conclusion, our study not only advances our understanding of the intricate interplay between hardware noise and GQML models but also lays the groundwork for informed strategies to enhance their resilience. As we navigate the challenges posed by noise in QML, our findings open new avenues for further exploration and optimization, offering a promising trajectory for the future development of robust and scalable GQML on quantum hardware.

\begin{acknowledgments}
CT is supported in part by the Helmholtz Association -``Innopool Project Variational Quantum Computer Simulations (VQCS)''. SC is supported by the quantum computing for earth observation (QC4EO) initiative of ESA $\Phi$-lab, partially funded under contract 4000135723/21/I-DT-lr, in the FutureEO program. SC, SV and MG are supported by CERN through the CERN Quantum Technology Initiative. This work is supported with funds from the Ministry of Science, Research and Culture of the State of Brandenburg within the Centre for Quantum Technologies and Applications (CQTA). This work is funded within the framework of QUEST by the European Union’s Horizon Europe Framework Programme (HORIZON) under the ERA Chair scheme with grant agreement No.\ 101087126. Access to the IBM Quantum Services was obtained through the IBM Quantum Innovation Centers at CERN and at DESY CQTA. Authors would like to thank Stefan K\"uhn, Tobias Hartung and Marco Cerezo for fruitful discussions. The views expressed here are those of the authors and do not reflect the official policy or position of IBM or the IBM~Quantum team. 
\begin{figure}[htb]
    \centering
    \includegraphics[width=0.1\textwidth]{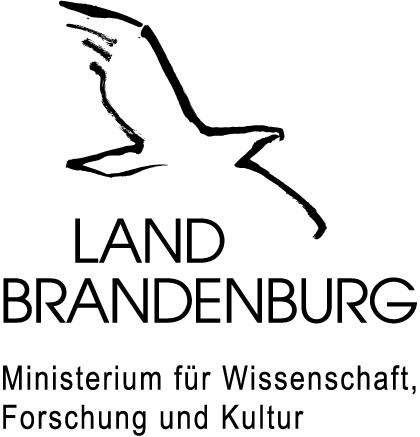}
\end{figure}
\end{acknowledgments}

\bibliography{main}

\onecolumngrid
\appendix

\section{Noise models}
\label{app:noise}

\subsection{Amplitude Damping Channel}

In Section~\ref{sec:noise}, we introduced the \textit{amplitude damping} (AD) channel with the following Kraus operators,

\begin{equation}
K_0 = \begin{bmatrix}
    1& 0\\
    0& \sqrt{1-\gamma}
\end{bmatrix}, \,
K_1 = \begin{bmatrix}
    0& \sqrt{\gamma}\\
    0& 0
\end{bmatrix}.
\end{equation}

These were given as matrices. Here, we also give them in the Pauli basis,

\begin{equation}
    K_0 = \frac{1+\sqrt{1-\gamma}}{2} \mbox{I} + \frac{1-\sqrt{1-\gamma}}{2} \mbox{Z}\, ,
    K_1 = \frac{\sqrt{\gamma}}{2} \mbox{X} - i\frac{\sqrt{\gamma}}{2} \mbox{Y}.
\end{equation}

This allows us to see the commutation of the AD channel with the $Z$ gate.

\subsection{Pauli Transfer Matrix formalism}
\label{app:ptm}

Working with the Kraus operators can become messy very quickly. \textit{Pauli transfer matrix} (PTM) formalism allows us to simplify this process~\cite{ptm}. In this formalism, we start by choosing the normalized Pauli basis $\hat{\mathbb{P}} = \frac{1}{\sqrt{2}} \{ I, X, Y, Z \}$. Then, the n-qubit operator $\hat{P} \in \hat{\mathbb{P}}^{\otimes n}$ can be represented as a basis vector $|P\rrangle \in \mathbb{R}^{4^n}$. 

We can also write the density matrix of a quantum state using this formalism. Consider the state $\ket{\psi}=\ket{0}$, which has the density matrix $\rho=\ket{\psi}\bra{\psi}=|0\rangle\langle0|$. The density matrix $\rho$ can be simply written as $\left[ 1/2, 0, 0, 1/2\right]$. This can easily be seen when $|0\rangle\langle0|$ is explicitly written as $(I+Z)/2$. 

Following this, a quantum channel $\EC \in \mathbb{R}^{4^n \times 4^n}$ becomes a matrix. Finally, the expectation value of the operator on the density matrix is simply $tr(\rho \hat{P})$. Then, using the PTM formalism we can compute the adjoint action of the unitaries as well as the noise channels as simple matrix multiplications.

Now, let's recall the Kraus operators of the Pauli channel $\NC_{\mbox{P}}$ are given as $K_0 = \sqrt{1-p_x - p_y - p_z}~I$, $K_1 = \sqrt{p_x}~X$, $K_2 = \sqrt{p_y}~Y$, $K_3 = \sqrt{p_z}~Z$. To obtain the PTM matrix of the Pauli channel we can write the action of the channel on all Pauli operators and perform state tomography. This will be fairly simple in this case.

\begin{align}
    \NC_{\mbox{P}} (I) &= I \\
    \NC_{\mbox{P}} (X) &= 1-2(p_y+p_z) X \\
    \NC_{\mbox{P}} (Y) &= 1-2(p_x+p_y) Y \\
    \NC_{\mbox{P}} (Z) &= 1-2(p_x+p_y) Z 
\end{align}

We will define the Pauli fidelity $f_P$ of a Pauli operator $P$ as the coefficient we observe in front (\textit{e.g.} $f_x = 1-2(p_y+p_z) $). Then, the PTM of the Pauli channel becomes, 

\begin{equation}
\Lambda_{\mbox{P}} = \begin{bmatrix}
    1 & 0 & 0 & 0\\
    0 & f_x & 0 & 0\\
    0 & 0 & f_y & 0\\
    0 & 0 & 0 & f_z
\end{bmatrix}.
\end{equation}

Following this, we can recover the \textit{bit flip} (BF), \textit{phase flip} (PF), \textit{depolarizing} (DP) channels' Kraus operators and the corresponding PTMs.

BF channel with probability $p$ becomes $K_0 = \sqrt{1-p}~I$, $K_1 = \sqrt{p}~X$. Then its PTM reads,

\begin{equation}
\Lambda_{\mbox{BF}} = \begin{bmatrix}
    1 & 0 & 0 & 0\\
    0 & 1 & 0 & 0\\
    0 & 0 & 1-2p & 0\\
    0 & 0 & 0 & 1-2p
\end{bmatrix}.
\end{equation}

PF channel with probability $p$ becomes $K_0 = \sqrt{1-p}~I$, $K_1 = \sqrt{p}~Z$. Then its PTM reads,

\begin{equation}
\Lambda_{\mbox{PF}} = \begin{bmatrix}
    1 & 0 & 0 & 0\\
    0 & 1-2p & 0 & 0\\
    0 & 0 & 1-2p & 0\\
    0 & 0 & 0 & 1
\end{bmatrix}.
\end{equation}

DP channel with probability $p$ becomes $K_0 = \sqrt{1-p}~I$, $K_1 = \sqrt{p/3}~X$, $K_2 = \sqrt{p/3}~Y$, $K_3 = \sqrt{p/3}~Z$. Then its PTM reads,

\begin{equation}
\Lambda_{\mbox{DP}} = \begin{bmatrix}
    1 & 0 & 0 & 0\\
    0 & 1-2p/3 & 0 & 0\\
    0 & 0 & 1-2p/3 & 0\\
    0 & 0 & 0 & 1-2p/3
\end{bmatrix}.
\end{equation}

PTM of the AD channel can also be obtained following the same procedure. Here we will skip this step and directly give the matrix.

\begin{equation}
\Lambda_{\mbox{AD}} = \begin{bmatrix}
    1 & 0 & 0 & 0\\
    0 & \sqrt{1-\gamma} & 0 & 0\\
    0 & 0 & \sqrt{1-\gamma} & 0\\
    \gamma & 0 & 0 & 1-\gamma
\end{bmatrix}
\end{equation}

Finally, we can use the PTM formalism to show the commutation of the Pauli Z with the AD channel. Recall that we need to satisfy the following for the commutation,

\begin{equation}
    \NC_{\mbox{AD}} \circ \mbox{Ad}_{Z} (\cdot) = \mbox{Ad}_{Z} \circ \NC_{\mbox{AD}} (\cdot)
\end{equation}

Then, it's easy to show this using the PTM formalism, 

\begin{equation}
\begin{bmatrix}
    1 & 0 & 0 & 0\\
    0 & \sqrt{1-\gamma} & 0 & 0\\
    0 & 0 & \sqrt{1-\gamma} & 0\\
    \gamma & 0 & 0 & 1-\gamma
\end{bmatrix} \cdot 
\begin{bmatrix}
    1 & 0 & 0 & 0\\
    0 & -1 & 0 & 0\\
    0 & 0 & -1 & 0\\
    0 & 0 & 0 & 1
\end{bmatrix} = 
\begin{bmatrix}
    1 & 0 & 0 & 0\\
    0 & -1 & 0 & 0\\
    0 & 0 & -1 & 0\\
    0 & 0 & 0 & 1
\end{bmatrix} \cdot
\begin{bmatrix}
    1 & 0 & 0 & 0\\
    0 & \sqrt{1-\gamma} & 0 & 0\\
    0 & 0 & \sqrt{1-\gamma} & 0\\
    \gamma & 0 & 0 & 1-\gamma
\end{bmatrix}.
\end{equation}

Since we are considering local noise models, the PTM of the $n$-qubit AD channel can be obtained by taking $n^{th}$ Kronecker power of the single qubit $\Lambda_{\mbox{AD}}$ \textit{i.e.} it is $\Lambda_{\mbox{AD}}^{\otimes n}$. Similarly, this also applies to $\mbox{Ad}_{Z}(\cdot)$, and as a result, we can conclude that $Z^{\otimes n}$ commutes with the $n$-qubit AD channel.

\section{Calculations for the toy model}
\label{app:toy}

In this section, we will give the details for the calculations in Section~\ref{sec:toy}. Let's start by recalling the definition of the toy model, which was described in Fig.~\ref{fig:1q-toy}. The data is encoded using the $R_Y$ gate and the redundant computation of $U U^\dagger$ is repeated $d$ times. The input state is chosen to be $\ket{+}$ The noise is modeled by applying the noisy operation between each $U$ and $U^\dagger$ gates. For simplicity, $U$ is chosen to be $R_Y(\theta)$, and the output of the model is considered to be the expectation value of the Pauli $X$. Then the final state of the model, before measurement, for input data $x$ is given as,

\begin{equation}
    \rho =  \mbox{Ad}_{R_Y(-\theta)} \circ \NC \circ \mbox{Ad}_{R_Y(\theta)} \circ \mbox{Ad}_{R_Y(-\theta)} \circ \cdots \circ \mbox{Ad}_{R_Y(\theta)} \circ  \mbox{Ad}_{R_Y(-\theta)} \circ \NC \circ \mbox{Ad}_{R_Y(\theta)} \circ \mbox{Ad}_{R_Y(x)} (|+\rangle\langle+|).
\end{equation}

The terms $\mbox{Ad}_{R_Y(\theta)}$ and $\mbox{Ad}_{R_Y(-\theta)}$ that appear next to each other will be identity. Then, this reduces to,

\begin{equation}
    \rho =  \mbox{Ad}_{R_Y(-\theta)} \circ \underbrace{\NC \circ \cdots \NC}_\text{d times} \circ \mbox{Ad}_{R_Y(\theta)} \circ \mbox{Ad}_{R_Y(x)} (|+\rangle\langle+|).
\end{equation}

We can compute this using the PTM of these terms. We already defined the PTM of the noise channels in Appendix~\ref{app:ptm}. Then, we give the definitions for the remaining terms here. The density matrix of $|+\rangle\langle+|$ can be written as, $(I+X)/2$. Then, it can be expressed with the vector $\left[1/2, 1/2, 0, 0\right]$. The PTM that represents the adjoint action of the $R_Y(\theta)$ gate can be expressed as,

\begin{equation}
    \mbox{Ad}_{R_Y(\theta)} = \begin{bmatrix}
    1 & 0 & 0 & 0\\
    0 & \cos(\theta) & 0 & -\sin(\theta)\\
    0 & 0 & 1 & 0\\
    0 & \sin(\theta) & 0 & \cos(\theta)
\end{bmatrix}.
\end{equation}

Furthermore, we need to point to the fact that the repetitive application of the noise channel will appear as the $d^{th}$ power of the PTM matrix of the corresponding noise channel. Finally, the expectation value of $X$ in the PTM picture will correspond to a dot product of the vector $\left[0, 1, 0, 0\right]$ with the final state. Then, let us write the full expression to obtain the expectation value under the Pauli channel, as it was given in Eq.~\ref{eq:toy-pauli},

\begin{equation}
    \hat{y}(x) = \begin{bmatrix}
    0 \\
    1 \\
    0 \\
    0
\end{bmatrix} \cdot
\begin{bmatrix}
    1 & 0 & 0 & 0\\
    0 & \cos(\theta) & 0 & \sin(\theta)\\
    0 & 0 & 1 & 0\\
    0 & -\sin(\theta) & 0 & \cos(\theta)
\end{bmatrix} \cdot
\begin{bmatrix}
    1 & 0 & 0 & 0\\
    0 & f_x^d & 0 & 0\\
    0 & 0 & f_y^d & 0\\
    0 & 0 & 0 & f_z^d
\end{bmatrix} \cdot
\begin{bmatrix}
    1 & 0 & 0 & 0\\
    0 & \cos(\theta) & 0 & -\sin(\theta)\\
    0 & 0 & 1 & 0\\
    0 & \sin(\theta) & 0 & \cos(\theta)
\end{bmatrix} \cdot
\begin{bmatrix}
    1 & 0 & 0 & 0\\
    0 & \cos(x) & 0 & -\sin(x)\\
    0 & 0 & 1 & 0\\
    0 & \sin(x) & 0 & \cos(x)
\end{bmatrix} \cdot
\begin{bmatrix}
    1/2 \\
    1/2 \\
    0 \\
    0
\end{bmatrix}.
\end{equation}

After the matrix multiplication, one obtains,

\begin{equation}
    \hat{y}(x) = ((f_x^d+f_z^d)\cos(x)+(f_x^d-f_y^d)\cos(x+2\theta))/4.
\end{equation}

Next, we would like to compute the output of the model under the AD channel. The PTM of the $d^{th}$ power of the AD channel results in a different structure, since it is not a diagonal matrix. This matrix can be given as follows, 

\begin{equation}
\Lambda_{\mbox{AD}}^{(d)} = \begin{bmatrix}
    1 & 0 & 0 & 0\\
    0 & (1-\gamma)^{d/2} & 0 & 0\\
    0 & 0 & (1-\gamma)^{d/2} & 0\\
    \Lambda_{\mbox{\tiny AD}_{(4,1)}}^{(d)} & 0 & 0 & (1-\gamma)^d
\end{bmatrix},
\end{equation}
where the term $\Lambda_{\mbox{\tiny AD}_{(4,1)}}^{(d)}$ corresponds to the matrix element of the index $(4,1)$. This term can be explicitly written as,

\begin{equation}
    \Lambda_{\mbox{\tiny AD}_{(4,1)}}^{(d)} \simeq d\gamma - \frac{d(d-1)}{2}\gamma^2.
\end{equation}

As also described in the main text, we skip writing the remaining terms as their contribution becomes negligible when realistic values are considered for the variables. For example, $\gamma \simeq 10^{-2}$ and $d < 20$. Then, this can be used to compute the expectation value under the AD channel. Using this, we can obtain the noisy prediction under the AD channel as,

\begin{align}
\hat{y}(x) &= \frac{1}{4}((1-\gamma)^{d}+(1-\gamma)^{d/2})\cos(x) \nonumber \\
        & + \frac{1}{4}((1-\gamma)^{d}-(1-\gamma)^{d/2})\cos(x+2\theta)\nonumber\\
        & + \frac{1}{2}\Lambda_{\mbox{\tiny AD}_{(4,1)}}^{(d)} \sin(\theta).
\end{align}

Next, we consider the combination of the Pauli channel with the AD channel. In the PTM formalism, their joint action can be represented as a matrix multiplication, such that $\Lambda_{\mbox{\tiny P+AD}} = \Lambda_{\mbox{\tiny P}} \cdot \Lambda_{\mbox{\tiny AD}}$ and it can be written as, 

\begin{equation}
\Lambda_{\mbox{\tiny P+AD}} = \begin{bmatrix}
    1& 0& 0& 0\\
    0& f_x\sqrt{1-\gamma}& 0& 0\\
    0& 0& f_y\sqrt{1-\gamma}& 0\\
    f_z\gamma& 0& 0& f_z(1-\gamma)
\end{bmatrix}.
\end{equation}

Then, $\Lambda_{\mbox{\tiny P+AD}}$ can be used to calculate the noisy predictions under the joint action of the Pauli and AD channels. This becomes, 

\begin{align}
\hat{y}(x) &= \frac{1}{4}(f_x^{d}(1-\gamma)^{d/2}+f_z^{d}(1-\gamma)^{d})\cos(x)\nonumber \\
        & + \frac{1}{4}(f_x^{d}(1-\gamma)^{d/2}-f_z^{d}(1-\gamma)^{d})\cos(x+2\theta) \nonumber\\
        & + \frac{1}{2}\Lambda_{\mbox{\tiny P+AD}_{(4,1)}}^{(d)} \sin(\theta).
\end{align}

and the $\Lambda_{\mbox{\tiny P+AD}_{(4,1)}}^{(d)}$ is,

\begin{equation}
    \Lambda_{\mbox{\tiny P+AD}_{(4,1)}}^{(d)} \simeq \left(  \sum_{k=1}^d f_z^{k}  \right)\gamma - \left(  \sum_{k=1}^d (k-1) \times f_z^{k}  \right)\gamma^2.
\end{equation}

\section{Impact of shot noise}

\begin{figure}[!h]
    \centering
    \includegraphics[width=\linewidth]{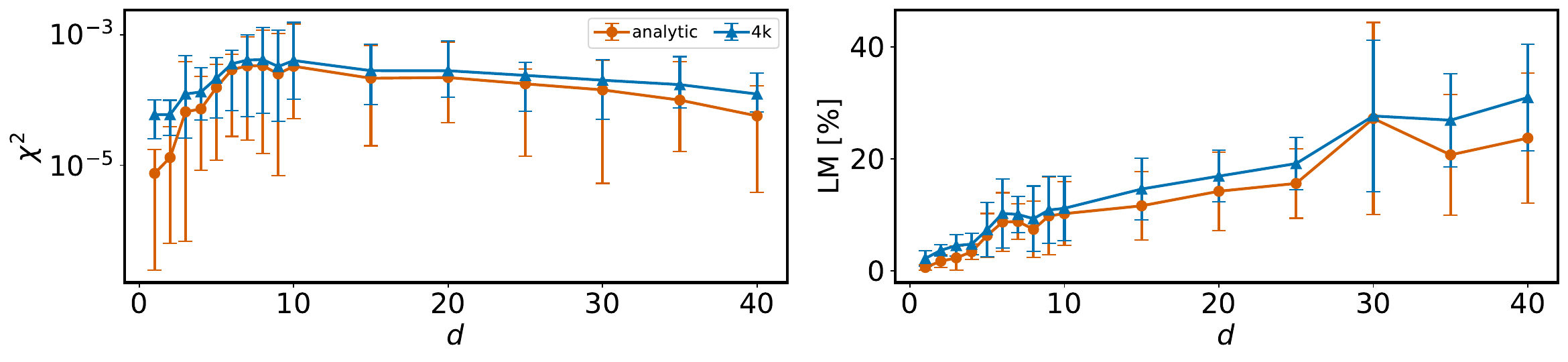}
    \caption{Comparison of symmetry breaking measurements with (4000 shots) and without shot noise. EQNN-Z model is simulated under AD channel with $\gamma = 0.01$.}
    \label{fig:shots}
\end{figure}

All simulations in the main text of the manuscript are performed with analytic expectation values omitting shot noise. All of the hardware runs are performed with 4000 shots. Here in Fig.~\ref{fig:shots}, we present the simulation of the EQNN-Z model simulated with AD channel using noise strength $\gamma = 0.01$ to show that the number of shots chosen is enough to match analytic results with high confidence.

\section{Zero noise extrapolation}
\label{app:zne}

Zero noise extrapolation (ZNE) is an error mitigation method that uses the expectation values measured at different noise strengths~\cite{kandala_error_2019}. These values can be extrapolated to zero noise level using Richardson's extrapolation method to obtain \textit{noiseless} expectation values. 

\begin{figure}[!h]
\subfloat[Analytical expectation values]{\includegraphics[width = 0.46\textwidth]
{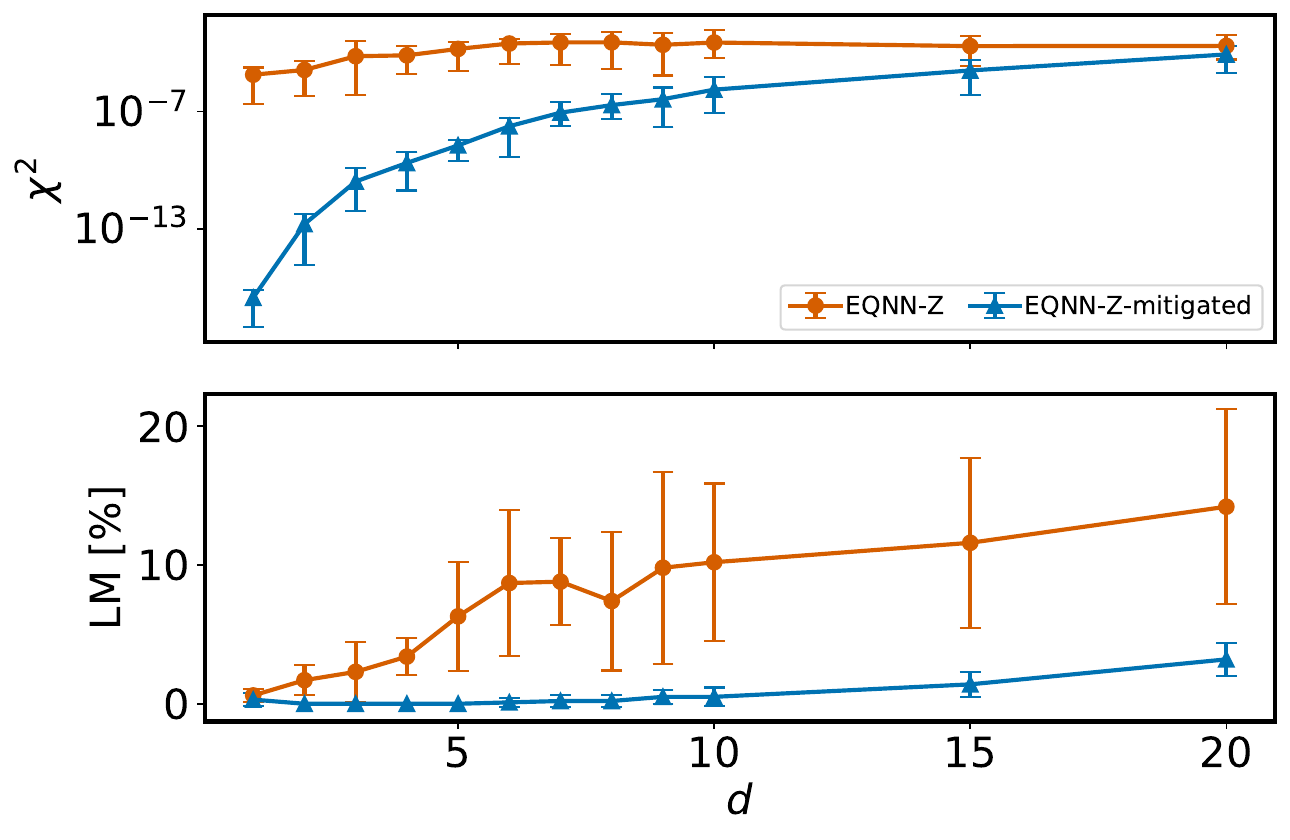}}
\hspace*{\fill}
\subfloat[Expectation values with 4k shots]{\includegraphics[width = 0.46\textwidth]
{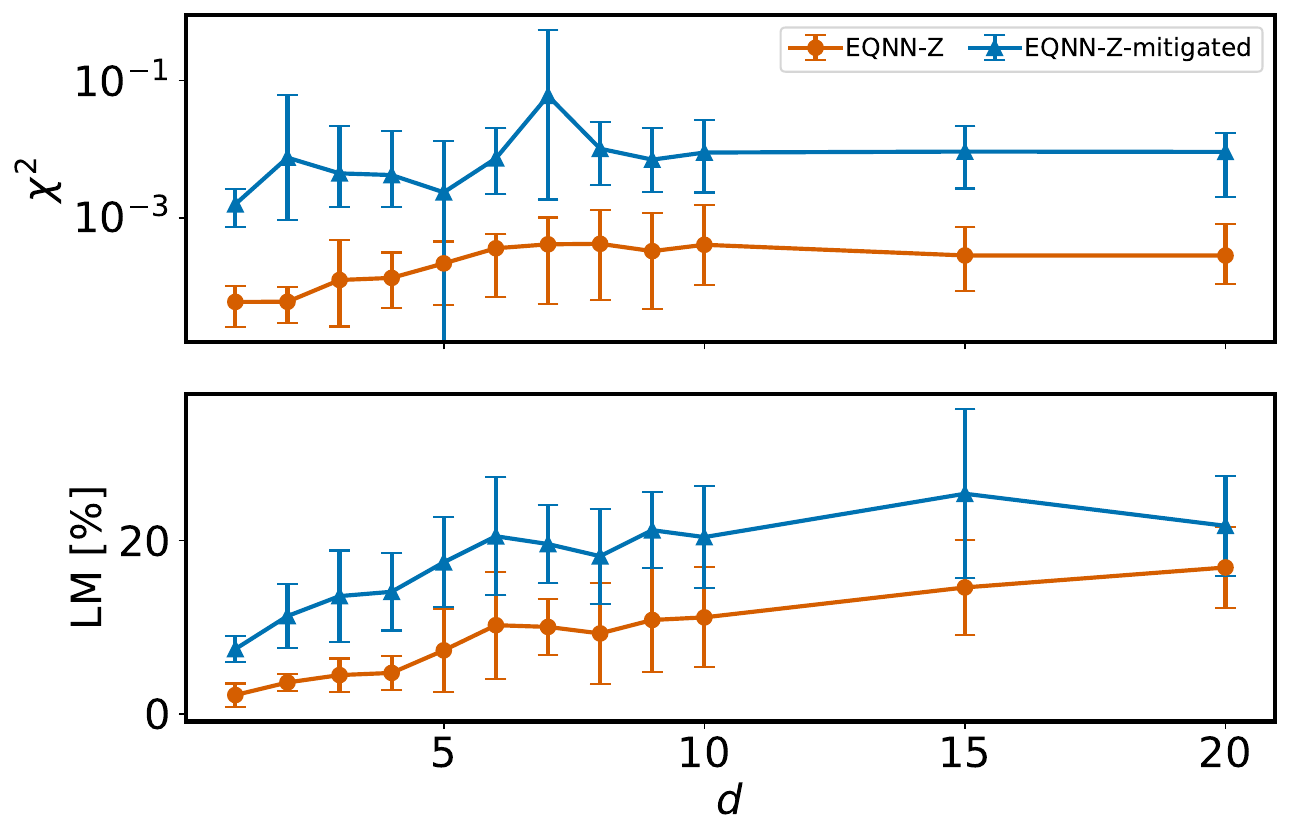}}
\caption{Symmetry breaking experiments with ZNE. EQNN-Z model used in Section~\ref{sec:2q-symm} is employed with and without shot noise.}
\label{fig:zne}
\end{figure}

We perform two separate numerical experiments to compare the effectiveness of ZNE. In the first one, the expectation values are computed analytically, while in the other one, the expectation values are computed using 4000 shots. In both experiments, the base noise level ($\lambda=1$) is chosen to be $\gamma = 0.01$. Then, the experiments are repeated using increasing levels of $\gamma \in \{ 0.015, 0.020, 0.025, 0.030\}$. These five expectation values for all noise scale factors are then extrapolated using Richardson's extrapolation to obtain the \textit{noiseless} expectation values. Results for varying number of layers in the presence of AD channel noise are presented in Fig.~\ref{fig:zne}. 

It is clear that ZNE can improve the accuracy of the results and bring LM values down significantly in the analytical case. However, when the number of shots is limited, ZNE fails and even worsens the results. This highlights the fact that ZNE requires many shots to work properly and the number of shots required will inevitably grow exponentially in the number of layers due to noise-induced BPs.

\section{\label{sec:hw} Hardware experiments}

In this section, we give details of the hardware experiments. All experiments are performed with the same settings using 4000 shots and no error mitigation method is used. The \textit{light optimization} is used to transpile the circuits, which includes the \textit{SABRE} method~\cite{li2019tackling}, 1Q gate optimization, and dynamical decoupling~\cite{viola1998dynamical}. The list of the devices, along with some of their properties, is presented in Table~\ref{tab:hw}.

\begin{table}[!h]
    \centering
    \begin{tabular}{|c|c|c|c|c|}
        \hline
        Name & T1 [us] & T2 [us] &  Gate time [ns] & Readout length [ns]\\
        \hline
        \cairo & 91.99 &  92.4 & 321.778 & 732.444 \\
        \hline
        \cusco & 126.78 & 78.77 & 460 & 4000 \\
        \hline
    \end{tabular}
    \caption{Properties of the physical quantum hardware used in this work. All values are reported as the median across all qubits on the chip. The values may change daily with each calibration.}
    \label{tab:hw}
\end{table}

\subsection{\label{sec:top} Hardware topology}

\begin{figure}[!h]
    \centering
    \includegraphics[width = 0.4\linewidth]{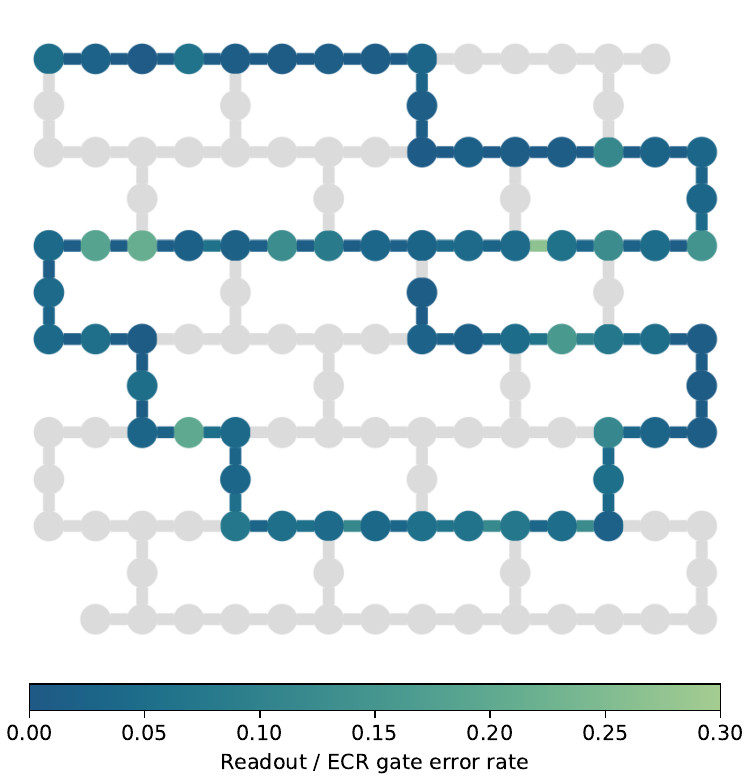}
    \caption{\label{fig:coupling_map} Coupling map of \cusco~and the qubit configuration chosen to run the quantum circuit of 64 qubits. The colors represent the readout error for each qubit and the two-qubit ECR gate for each qubit connection. }
\end{figure}

The coupling map of \cusco~used for the 64 qubit experiments is presented in Fig.~\ref{fig:coupling_map}. We choose a suitable nearest neighbor set of qubits to have 1D connectivity. 

\subsection{\label{sec:pulse} Pulse efficient transpilation}

In order to run a generic quantum circuit on the real IBM Quantum hardware, the circuit should first transpired into the set of basis gates, which are pre-calibrated on the corresponding hardware. The automatic IBM quantum transpilation only exposes fixed-frequency cross-resonance gates~\cite{chow2011simple}. 
Instead of the fixed frequency gates, we can use the continuous gate native to the quantum hardware. For the low rotation angles, the circuit duration becomes shorter, leading to less decoherence noise and more accurate results.

\begin{figure}[!h]
\centering

\begin{minipage}{0.7\linewidth}
\begin{tcolorbox}

\begin{verbatim}
rzx_basis = ['rzx', 'rz', 'x', 'sx']

pm = PassManager([
    # Consolidate consecutive two-qubit operations.
    Collect2qBlocks(),
    ConsolidateBlocks(basis_gates=['rz', 'sx', 'x', 'rxx']),
 
    # Rewrite circuit in terms of Weyl-decomposed echoed RZX gates.
    EchoRZXWeylDecomposition(backend),
 
    # Attach scaled CR pulse schedules to the RZX gates.
    RZXCalibrationBuilderNoEcho(backend),
 
    # Simplify single-qubit gates.
    UnrollCustomDefinitions(std_eqlib, rzx_basis),
    BasisTranslator(std_eqlib, rzx_basis),
    Optimize1qGatesDecomposition(rzx_basis),
]) 
\end{verbatim}
\end{tcolorbox}

\end{minipage}
\caption{\label{fig:implementation}Python code for RZX transpilation in Qiskit implementation taken from Ref.~\cite{qiskit_release}.}
\end{figure}

The calibrated CNOT gates are built with a GuassianSquare pulse, which is a flat-top pulse with the area, 
\begin{equation}
    \alpha* = \norm{A*}[w* + \sqrt{2\pi}\sigma \cdot  erf(\frac{rf}{\sqrt{2} \sigma}) ], 
\end{equation}
with $A*$ the amplitude, $w*$ the width, $rf$ the risefall and $\sigma$ the standard deviation of the corresponding Gaussian flanks~\cite{Earnest2021}. 
The pulse of $RZX(\theta)$ gate is created by rescaling the area $\alpha$ as~\cite{stenger2021simulating}
\begin{equation}
    \alpha(\theta) = \frac{2\theta \alpha* }{\pi}.  
\end{equation}

In the Qiskit implementation, the RZX-based transpilation works as shown in Fig.~\ref{fig:implementation}.  
First of all, we collect all the consecutive two-qubit operations and consolidate them into a general two-qubit $SU(4)$ operation. 
Then, the corresponding two-qubit gate is decomposed in terms of echoed RZX gates by leveraging Cartan's decomposition~\cite{khaneja2001cartan}. Those gates are calibrated by scaling the Gaussian square pulses of the fixed-frequency CNOT or ECR gates.  
Finally, the single-qubit gates are simplified and optimized. 

\begin{figure}[!h]
    \centering
    \hspace*{\fill}
    \subfloat[ECR decomposition]{\includegraphics[width=0.55\linewidth]{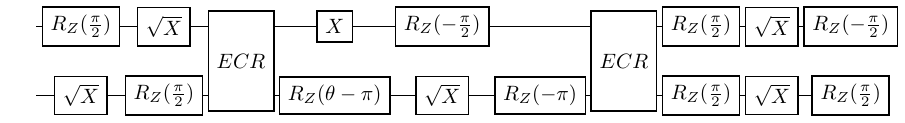}} 
    \hspace*{\fill}
    \newline
    \subfloat[RZX-based decomposition with the built-in echo. ]{\includegraphics[width=0.65\linewidth]{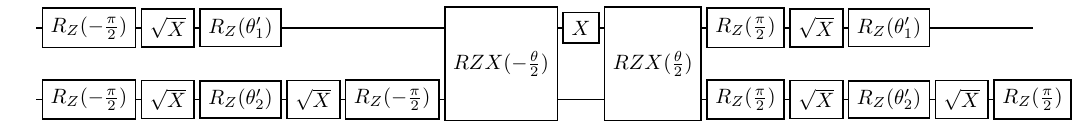}}
    
    \caption{Circuit decomposition of $RXX(\theta)$ gate on \cusco. $\theta'_1$ and $\theta'_2$ in (b) are the single-qubit rotation angles computed by Cartan's decomposition  }
    \label{fig:circuit_decomposition}
\end{figure}

\begin{figure}[!h]
\subfloat[\label{fig:pulse_schedule_ecr} Pulse schedule for ECR decomposition]{\includegraphics[width = 0.46\textwidth]
{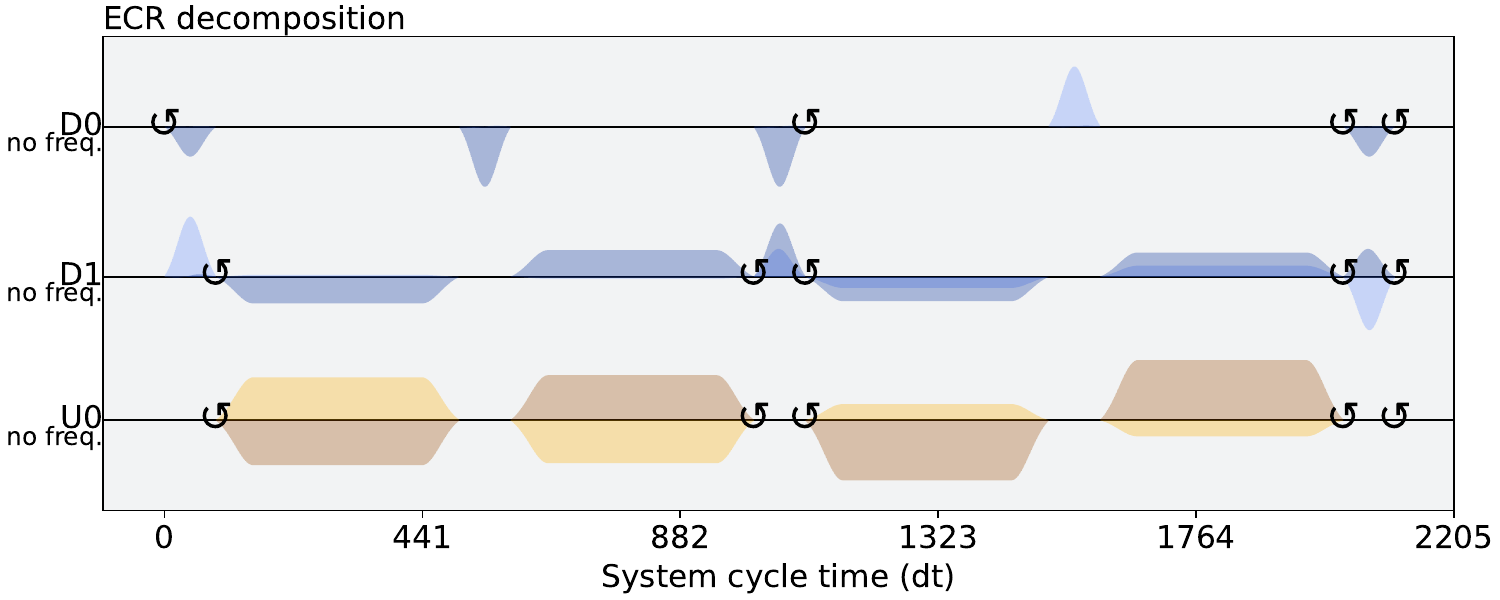}}
\hspace*{\fill}
\subfloat[\label{fig:pulse_schedule_rzx} Pulse schedule for RZX decomposition]{\includegraphics[width = 0.46\textwidth]
{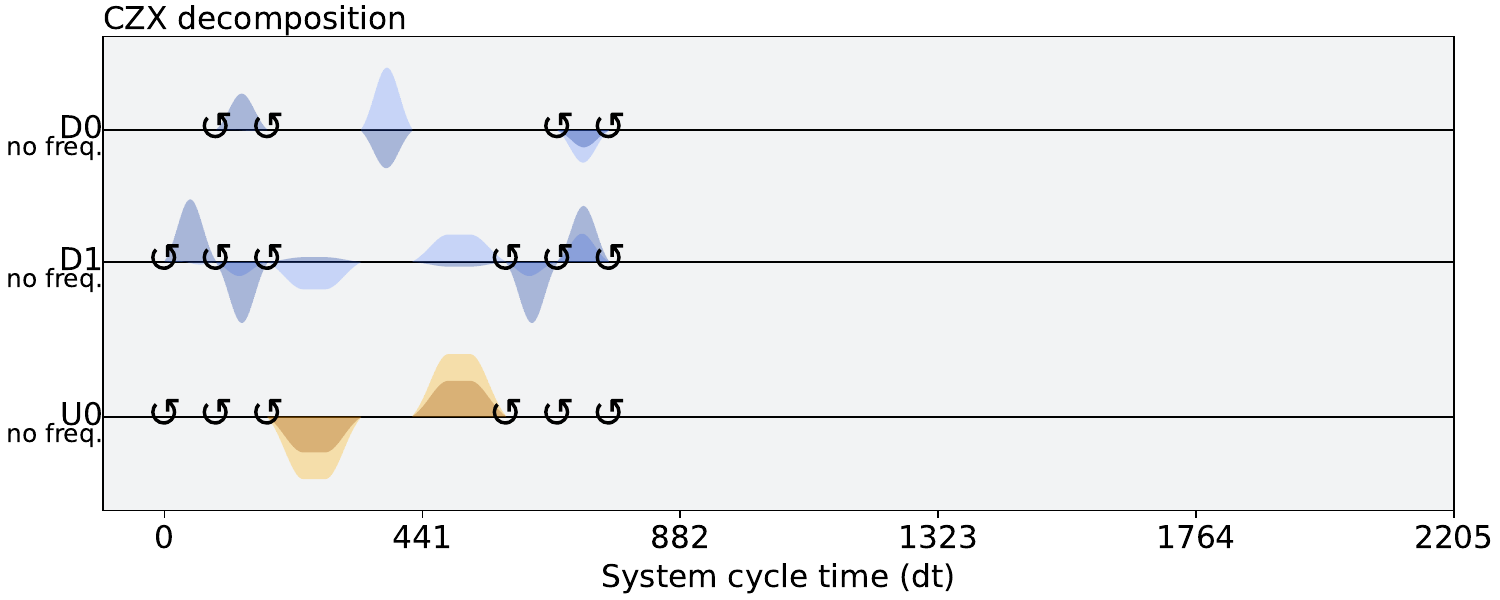}}
    \caption{\label{fig:pulse_schedule} Pulse schedule for ECR-based decomposition and pulse-efficient RZX-based decompositions (c.f. Fig.~\ref{fig:circuit_decomposition}). The symbol $\circlearrowleft$ indicates the virtual $Z$ gates~\cite{mckay2017efficient}.  }
\end{figure}

Fig.~\ref{fig:circuit_decomposition} and Fig.~\ref{fig:pulse_schedule} display the decomposed circuits of the RXX gate for ECR-based and RZX-based decomposition using the basis gates on \cusco~and the corresponding pulse schedule, respectively.  As shown in Fig.~\ref{fig:pulse_schedule}, the pulse schedule with RZX decomposition is much shorter compared to the one with ECR decomposition, resulting in less decoherence and better results, as mentioned previously.

\end{document}